\documentclass[epj]{svjour}
\usepackage{graphics}
\begin{document}
\title{Triton photodisintegration in three-dimensional approach}
\author{M. A. Shalchi \and S. Bayegan
}                     

\institute{Department of Physics, University of Tehran, P.O.Box
14395-547, Tehran, Iran\\ \email{shalchi@khayam.ut.ac.ir}}
\date{Received: date / Revised version: date}
%
\abstract{ Two- and three- particles photodisintegration of the
triton is investigated in a three-dimensional (3D) Faddeev approach.
For this purpose the Jacobi momentum vectors for three particles
system and spin-isospin quantum numbers of the individual nucleons
are considered. Based on this picture the three-nucleon Faddeev
integral equations with the two-nucleon interaction are formulated
without employing the partial wave decomposition. The single nucleon
current as well as $\pi-$ and $\rho-$ like exchange currents are
used in an appropriate form to be employed in 3D approach. The
exchange currents are derived from AV18 NN force. The two-body
t-matrix, Deuteron and Triton wave functions are calculated in the
3D approach by using AV18 potential. Benchmarks are presented to
compare the total cross section for the two- and three- particles
photodisintegration in the range of $E_{\gamma}<30~ MeV$. The 3D
Faddeev approach shows promising results.
} 
\maketitle
\section{Introduction}
\label{sec:introduction} The study of 3N system with the
electromagnetic interaction shed light on the ambiguity of different
kinds of nuclear forces participating in the few body systems.
Following the introduction of Faddeev formulation for the three-body
system \cite{Faddeev_ThF39}-\cite{Alt_NPB2}, a new effort using this
scheme was started. The electrodisintegration \cite{Lehman_PhRL23}
and photodisintegration \cite{Barbour_PhRL19} of $^3\!$He and
$^3\!$H in the Faddeev scheme were performed as early attempts in
this respect. The photodisintegration calculation based on Faddeev
formalism for 3N bound and continuum with the same 3N Hamiltonian
were performed with quite simple separable NN interactions
\cite{Gibson_PhRC11}. The other approaches in this field can be
considered to be
 Green Function Monte-Carlo \cite{Carlson_PhRC36} and
Lorentz Integral Transform \cite{Efros_PhLB338} methods.

The photodisintegration of $^3\!$He was investigated by
pair-correlated hyperspherical harmonics method using \\AV18 NN and
Urbana 3N forces \cite{Viviani_PhRC61}. Also, a systematic
application of Faddeev formalism was introduced, based on the
partial wave decomposition, for the calculation of the
photodisintegration of $^3\!$He and $^3\!$H with the two- and three-
body forces \cite{golak}.

We use the novel three-dimensional (3D) approach
\cite{rice,Fachruddin_PhRC62,Fachruddin_PhRC63,hadizadeh-Eur.Phys,Bayegan-Prog.Theor.Phys,Bayegan-nucl814,Bayegan_PhRC77,Bayegan_PhRC79,Bayegan-nucl832,harzchi-nucla46,Hadizadeh-Phys.
Revc83}. In this approach we employ the momentum-vector variables as
basis states. We derive the Faddeev integral equation in a realistic
3D scheme as a function of Jacobi momentum vectors and the
spin-isospin quantum numbers. We implement this 3D scheme in the
two- (Nd) and three-body (3N) photodisintegration of $^3$H. This 3D
approach avoids truncation problems and the necessity of complicated
recoupling algebra that accompanies partial wave based calculations.

It is our aim to calculate the nuclear matrix element which is given
by $N^{3N}=\frac{1}{2}\,^{(-)}\langle\phi_0|P|U\rangle$ or
\\$N^{Nd}=\frac{1}{2}\langle\phi_d|P|U\rangle$ for
$\gamma\,^3H\rightarrow 3N$ or $\gamma\,^3H\rightarrow Nd$,
respectively. We introduce $^{(-)}\langle\phi_0|$ and
$\langle\phi_d|$ as outgoing 3N and Nd system, $P$ as permutation
operator and $|U\rangle=(1+P)O|\Psi\rangle+(tG_0P+...)|U\rangle$ as
an auxiliary state with $O$ as the current operator. The wave
function $\Psi$ is the three-body bound state, $t$ is the two body
transition operator and $G_0$ is free propagator. The second part of
$|U\rangle$ is related to the solution of $|U\rangle$ of this
equation. We introduce a formalism to solve this equation directly
in the momentum-vector variables avoiding the partial wave (PW)
decomposition scheme.

We use the AV18 potential for the calculation of wave functions of
the triton and deuteron as well as the two-body t-matrix. For
nuclear electromagnetic current operator we choose, in addition to
the single-nucleon current, the two-body contribution in the form of
the $\pi-$ and $\rho-$ meson  exchanges with connection to NN force
AV18.

This manuscript is organized as follow: In section
\ref{sec:formulation} formulation of nuclear matrix equation in the
3D approach is introduced and $|U\rangle$, as an auxiliary state, is
derived for the two-body interaction with handling of singularity
problem. In this section the appropriate formulation is constructed
for the one- and two-body currents to be implemented in the 3D
formalism. In section \ref{sec:numerical} numerical method and
results are presented and section \ref{sec:summary} is concluded by
a summary and outlook.
\section{Formulation of N matrix equation in a 3D approach}
\label{sec:formulation} Our aim is to calculate the total cross
section of the triton photodisintegration. Final states can be
either Nd or free 3N state. For each we can write the cross section
as a function of the tensor components of the nuclear matrix
element, which is the nuclear current sandwiched between initial
bound and final scattering states:

\begin{eqnarray} \label{eq:1}
 \sigma_{t}^{Nd}=\int
&&d\hat{k}_n(2\pi)^4\alpha\frac{1}{2Q}\frac{k_n^2}{|\frac{k_n}{m_N}-\frac{\textbf{k}_d\cdot\textbf{k}_n}{2m_Nk_n}|}\nonumber
\\&&\times\frac{1}{2}\sum_{m,m_n,m_d}(|N_{+1}^{Nd}|^2+|N_{-1}^{Nd}|^2),
\end{eqnarray}
\begin{eqnarray} \label{eq:1p}
&&\sigma_{t}^{3N}=\int
d\hat{k}_1d\hat{k}_2dk_1\frac{1}{2}\sum_{m,m_1,m_2,m_3}\frac{2\pi^2}{Q}\alpha\nonumber
\\&&\frac{k_1^2k_2^2m_N}{|Q\cos\theta_2-k_1\cos\theta_{12}-2k_2|}(|N_{+1}^{3N}|^2+|N_{-1}^{3N}|^2).
\end{eqnarray}

In the above equations $m_N$ is the nucleon mass,
$\alpha\approx\frac{1}{137}$ is the fine-structure constant,
$\textbf{k}_n$ and $\textbf{k}_d$ are the neutron and deuteron
momentums, respectively. $\textbf{k}_i$s are free particle momenta
and $\theta_i$s are their angles with $\textbf{Q}$ as the momentum
of the photon. $\theta_{12}$ is the angle between momenta $k_1$ and
$k_2$. There is also a summation over all the spins of ingoing and
outgoing particles which are indicated by $m$, $m_d$, $m_1$, $m_2$
and $m_3$.

To obtain the nuclear matrix element we follow the formalism in
\cite{golak}.
\begin{eqnarray} \label{eq:2}
N^{3N}=\frac{1}{2}\,^{(-)}\langle\phi_0|P|U\rangle,
\end{eqnarray}
 \begin{eqnarray} \label{eq:2p}
N^{Nd}=\frac{1}{2}\langle\phi_d|P|U\rangle,
\end{eqnarray}
where $|\phi_0\rangle^{(-)}$ is the three particles scattering
 state as;
\begin{eqnarray} \label{eq:3}
|\phi_0\rangle^{(-)}=(1+G_0t)|\phi_0\rangle,
\end{eqnarray}
so we can write:
\begin{eqnarray}\label{eq:3.5}
N^{3N}=\frac{1}{2}[\langle\phi_0|P|U\rangle+\langle\phi_0|tG_0P|U\rangle]
\end{eqnarray}
 and $|\phi_d\rangle$ is Nd state
\begin{eqnarray} \label{eq:4}
|\phi_d\rangle=|\textbf{q}m_1\nu_1\rangle|\psi_d\rangle,
\end{eqnarray}
 where $\psi_d$ is the deuteron wave function, and $|U\rangle$ is the auxiliary state that can be calculated by solving
 the following equation.
\begin{eqnarray} \label{eq:5}
|U\rangle=(1+P)J|\psi\rangle+tG_0P|U\rangle.
\end{eqnarray}
$|\psi\rangle$ is the three-body bound state, $J$ is the current,
$G_0$ is the 3N free propagator, $t$ is the two-body transition
operator and $P$ is the permutation operator.

\subsection{Introducing free basis states} \label{subsec:Faddeev-equations}

We introduce $|\phi_0\rangle$ as our free basis states where the
particles 2 and 3 are in subsystem and particle 1 is the spectator.
The state is antisymmetric under permutation of subsystem particles.
\begin{eqnarray} \label{eq:6}
|\phi_0\rangle&&=|\textbf{p}m_2m_3\nu_2\nu_3\rangle^a|
\textbf{q}m_1\nu_1\rangle\nonumber
\\&&=|\textbf{p}\textbf{q}m_1m_2m_3\nu_1\nu_2\nu_3\rangle^a,
\end{eqnarray}
where $m_i$ and $\nu_i$ are spin and isospin of the individual
nucleons and $\textbf{p}$ and $\textbf{q}$ are Jacobi momentum for a
three-particle system.
\begin{eqnarray} \label{eq:7}
&&\textbf{p}=\frac{1}{2}(\textbf{k}_2-\textbf{k}_3),\\
&&\textbf{q}=\frac{2}{3}[\textbf{k}_1-\frac{1}{2}(\textbf{k}_2+\textbf{k}_3)],
\end{eqnarray}
where $\textbf{k}_1$, $\textbf{k}_2$, $\textbf{k}_3$ are momentums
of the particles. So we have:
\begin{eqnarray} \label{eq:8}
 &&|\textbf{p}\textbf{q}m_1m_2m_3\nu_1\nu_2\nu_3\rangle^a
 \nonumber \\&&=\frac{1}{2}(|\textbf{p}\textbf{q}m_1m_2m_3\nu_1\nu_2\nu_3\rangle\nonumber \\&&-|-
 \textbf{p}\textbf{q}m_1m_3m_2\nu_1\nu_3\nu_2\rangle).
\end{eqnarray}
\subsection{Nuclear matrix in free basis states} \label{subsec:Faddeev-equations}
We now apply these free basis states to the left hand side of each
term of the Eq. (\ref{eq:5}):
\begin{eqnarray} \label{eq:11}
\langle\phi_0|U\rangle&&=\,^a\langle\textbf{p}\textbf{q}m_1m_2m_3\nu_1\nu_2\nu_3|U\rangle
\nonumber
\\&&=\,^a\langle\textbf{p}\textbf{q}m_1m_2m_3\nu_1\nu_2\nu_3|(1+P)J|\psi\rangle\nonumber
\\&&+\,^a\langle\textbf{p}\textbf{q}m_1m_2m_3\nu_1\nu_2\nu_3|tG_0P|U\rangle.
\end{eqnarray}
In the next step we need to know the effect of the permutation
operator on the free basis states.
\begin{eqnarray} \label{eq:12}
&&P|\textbf{p}\textbf{q}m_1m_2m_3\nu_1\nu_2\nu_3\rangle^a \nonumber
\\&&=|(-\frac{1}{2}\textbf{p}-\frac{3}{4}\textbf{q})(\textbf{p}-\frac{1}{2}\textbf{q})m_2m_3m_1\nu_2\nu_3\nu_1\rangle^a\nonumber
\\&&+|(-\frac{1}{2}\textbf{p}+\frac{3}{4}\textbf{q})(-\textbf{p}-\frac{1}{2}\textbf{q})m_3m_1m_2\nu_3\nu_1\nu_2\rangle^a,
\end{eqnarray}
so the first part of Eq. (\ref{eq:3.5}) is
\begin{eqnarray} \label{eq:25}
&&\langle\phi_0|P|U\rangle=\,^a\langle\textbf{p}\textbf{q}m_1m_2m_3\nu_1\nu_2\nu_3|P|U\rangle\nonumber
\\&&=\,^a\langle(-\frac{1}{2}\textbf{p}-\frac{3}{4}\textbf{q}),(\textbf{p}-\frac{1}{2}\textbf{q})m_2m_3m_1\nu_2\nu_3\nu_1|U\rangle\nonumber
\\&&+\,^a\langle(-\frac{1}{2}\textbf{p}+\frac{3}{4}\textbf{q}),(-\textbf{p}-\frac{1}{2}\textbf{q})m_3m_1m_2\nu_3\nu_1\nu_2|U\rangle.
\end{eqnarray}
In the evaluation of the first term of Eq. (\ref{eq:11}) we can
write
\begin{eqnarray} \label{eq:13}
&&^a\langle\textbf{p}\textbf{q}m_1m_2m_3\nu_1\nu_2\nu_3|(1+P)J|\Psi\rangle\nonumber
\\&&=\,^a\langle\textbf{p}\textbf{q}m_1m_2m_3\nu_1\nu_2\nu_3|J|\Psi\rangle\nonumber
\\&&+\,^a\langle(-\frac{1}{2}\textbf{p}-\frac{3}{4}\textbf{q})(\textbf{p}-\frac{1}{2}\textbf{q})\,m_2m_3m_1\nu_2\nu_3\nu_1|J|\Psi\rangle\nonumber
\\&&+\,^a\langle(-\frac{1}{2}\textbf{p}+\frac{3}{4}\textbf{q})(-\textbf{p}-\frac{1}{2}\textbf{q})\,m_3m_1m_2\nu_3\nu_1\nu_2|J|\Psi\rangle.\;\;
\end{eqnarray}
Evaluation of each term of Eq. (\ref{eq:13}) is similar and
represents generally, by using completeness relation and inserting
the free basis states in a suitable position:
\begin{eqnarray} \label{eq:14}
&&^a\langle\textbf{p}\textbf{q}m_1m_2m_3\nu_1\nu_2\nu_3|J|\Psi\rangle=\sum_{m'\nu'}\int
d\textbf{p}'d\textbf{q}'\nonumber \\
&&\times\,^a\langle\textbf{p}\textbf{q}m_1m_2m_3\nu_1\nu_2\nu_3|J|\textbf{p}'\textbf{q}'m'_1m'_2m'_3\nu'_1\nu'_2\nu'_3\rangle^a\nonumber
\\&&\times\Psi_{m'_1m'_2m'_3}^{\nu'_1\nu'_2\nu'_3}
                                         (\textbf{p}',\textbf{q}'),
\end{eqnarray}
triton wave function in the free basis states is introduced
symbolically as follows:
\begin{eqnarray} \label{eq:15}
\Psi_{m_1m_2m_3}^{\nu_1\nu_2\nu_3}(\textbf{p},\textbf{q})=\,^a\langle\textbf{p}\textbf{q}m_1m_2m_3\nu_1\nu_2\nu_3|\Psi\rangle.
\end{eqnarray}

For the second part of Eq. (\ref{eq:11}) we have:
\begin{eqnarray} \label{eq:16}
&&\langle\phi_0|tG_0P|U\rangle=\,^a\langle\textbf{p}\textbf{q}m_1m_2m_3\nu_1\nu_2\nu_3|tG_0P|U\rangle\nonumber
\\&&=\sum_{m',\nu',m'',\nu''}\int d\textbf{p}'d\textbf{q}'d\textbf{p}''d\textbf{q}''
\nonumber
\\\,&&^a\langle\textbf{p}\textbf{q}m_1m_2m_3\nu_1\nu_2\nu_3|tG_0|\textbf{p}'\textbf{q}'m'_1m'_2m'_3\nu'_1\nu'_2\nu'_3\rangle^a\nonumber
\\
&&\times\,^a\langle\textbf{p}'\textbf{q}'m'_1m'_2m'_3\nu'_1\nu'_2\nu'_3|P|\textbf{p}''\textbf{q}''m''_1m''_2m''_3\nu''_1\nu''_2\nu''_3\rangle^a\nonumber
\\ &&\times
\,^a\langle\textbf{p}''\textbf{q}''m''_1m''_2m''_3\nu''_1\nu''_2\nu''_3|U\rangle,
\end{eqnarray}
there are two  important terms in Eq. (\ref{eq:16}) to be evaluated.
The first term including $tG_0$ operators can be written in terms of
two body t-matrices in the free basis-states multiplied by free
propagator, $G_0=\frac{1}{E-H_0}$, in the energy of subsystem. For
the second term of Eq. (\ref{eq:16}), one can proceed to the final
form by considering the properties of the permutation operator, $P$
( Eq. (\ref{eq:12})), and the symmetry properties of the free basis
states. The final form of the second part of Eq. (\ref{eq:11}) is
evaluated as follows:
\begin{eqnarray} \label{eq:22}
&&\langle\phi_0|tG_0p|U\rangle=\,^a\langle\textbf{p}\textbf{q}m_1m_2m_3\nu_1\nu_2\nu_3|tG_0P|U\rangle\nonumber
\\&&=2\sum_{m'_2m'_3\nu'_2\nu'_3}\int d\textbf{q}''
d\textbf{q}'\frac{1}{E-\frac{3q^2}{4m}-\frac{{\Pi_2}^2}{m}}\delta(\textbf{q}'-\textbf{q})\nonumber
\\
&&\times\,^a\langle\textbf{p}m_2m_3\nu_2\nu_3|t|-\mathbf{\Pi_2}m'_2m'_3\nu'_2\nu'_3\rangle^a\nonumber
\\\,
&&\times^a\langle\mathbf{\Pi_1}\textbf{q}''m'_3m_1m'_2\nu'_3\nu_1\nu'_2|U\rangle,
\end{eqnarray}

where the shifted argument $\mathbf{\Pi_1}$ and $\mathbf{\Pi_2}$ are
defined as follows
\begin{eqnarray} \label{eq:19}
\mathbf{\Pi_1}=\mathbf{q}'+\frac{1}{2}\textbf{q}'',~~~~~~~~~~~~~~~\mathbf{\Pi_2}=\mathbf{q}''+\frac{1}{2}\textbf{q}'.
\end{eqnarray}

We write for Eq. (\ref{eq:11}) with the help of Eq. (\ref{eq:22}):
\begin{eqnarray} \label{eq:26}
&&^a\langle\textbf{p}\textbf{q}m_1m_2m_3\nu_1\nu_2\nu_3|U\rangle\nonumber
\\&&=^a\langle
\textbf{p}\textbf{q}m_1m_2m_3\nu_1\nu_2\nu_3|(1+P)J|\psi\rangle\nonumber
\\&&+2\sum_{m'_2m'_3\nu'_2\nu'_3}\int
d\textbf{q}''\frac{1}{E-\frac{q^2+q''^2+\textbf{q}\cdot\,\textbf{q}''}{m}}\nonumber
\\
&&\times\,^a\langle\textbf{p}m_2m_3\nu_2\nu_3|t|-\textbf{q}''-\frac{1}{2}\textbf{q},m'_2m'_3\nu'_2\nu'_3\rangle^a\nonumber
\\
&&\times\,^a\langle\textbf{q}+\frac{1}{2}\textbf{q}'',\textbf{q}''m'_3m_1m'_2\nu'_3\nu_1\nu'_2|U\rangle,
\end{eqnarray}
where the first term has been described before in Eq. (\ref{eq:13}).

Finally the calculation of $N^{3N}$ after inserting Eqs.
(\ref{eq:25}) and (\ref{eq:22}) in Eq. (\ref{eq:3.5}) leads to:
\begin{eqnarray} \label{eq:23}
N^{3N}&&=\frac{1}{2}\{\,^a\langle(-\frac{1}{2}\textbf{p}-\frac{3}{4}\textbf{q}),(\textbf{p}-\frac{1}{2}\textbf{q})m_2m_3m_1\nu_2\nu_3\nu_1|U\rangle\nonumber
\\&&+\,^a\langle(-\frac{1}{2}\textbf{p}+\frac{3}{4}\textbf{q}),(-\textbf{p}-\frac{1}{2}\textbf{q})m_3m_1m_2\nu_3\nu_1\nu_2|U\rangle\}\nonumber
\\&&+\sum_{m'_2m'_3\nu'_2\nu'_3}\int
d\textbf{q}''\frac{1}{E-\frac{q^2+q''^2+\textbf{q}\cdot\,\textbf{q}''}{m}}\nonumber
\\
&&\times\,^a\langle\textbf{p}m_2m_3\nu_2\nu_3|t|-\textbf{q}''-\frac{1}{2}\textbf{q},m'_2m'_3\nu'_2\nu'_3\rangle^a\nonumber
\\\,
&&\times^a\langle\textbf{q}+\frac{1}{2}\textbf{q}'',\textbf{q}''m'_3m_1m'_2\nu'_3\nu_1\nu'_2|U\rangle,
\end{eqnarray}
we proceed to $N^{Nd}$ calculation by inserting the completeness
relation of the free basis states and Eq. (\ref{eq:25}) into:
\begin{eqnarray} \label{eq:24}
&&N^{Nd}=\frac{1}{2}\sum_{m',\nu'}\int
d\textbf{p}'d\textbf{q}'\langle\psi_d^{M_d}|\langle\textbf{q}m_1\nu_1|\textbf{p}'\textbf{q}'m'_1m'_2m'_3\nu'_1\nu'_2\nu'_3\rangle^a\nonumber
\\
\,&&\times^a\langle\textbf{p}'\textbf{q}'m'_1m'_2m'_3\nu'_1\nu'_2\nu'_3|P|U\rangle\nonumber\\
&&=\frac{1}{2}\sum_{m'_2m'_3\nu'_2\nu'_3}\int
d\textbf{p}'\langle\psi_d^{M_d}|\textbf{p}'m'_2m'_3\nu'_2\nu'_3\rangle^a\nonumber
\\&&\times[
^a\langle-\frac{1}{2}\textbf{p}'-\frac{3}{4}\textbf{q},\textbf{p}-\frac{1}{2}\textbf{q},m'_2m'_3m_1\nu'_2\nu'_3\nu_1|U\rangle\nonumber\\
&&+^a\langle-\frac{1}{2}\textbf{p}'+\frac{3}{4}\textbf{q},-\textbf{p}-\frac{1}{2}\textbf{q},m'_3m_1m'_2\nu'_3\nu_1\nu'_2|U\rangle].
\end{eqnarray}
\subsection{Singularity problem and rewriting Eq. (\ref{eq:26})} \label{subsec:Faddeev-equations}
We know that the two-body t-matrix has a simple singularity at
$E=E_d$, which is the deuteron binding energy. There is also a
moving singularity in Eq. (\ref{eq:22}), which is very difficult to
handle. To solve this moving singularity problem we use the method
introduced in Ref.\cite{Elster_FbS45} and we separate the radial
part from the angle part in the Dirac delta functions.
\begin{eqnarray} \label{eq:27}
&&\delta(\textbf{p}'+\mathbf{\Pi_2})\delta(\textbf{p}''-\mathbf{\Pi_1})\nonumber
\\&&=\frac{\delta(p'-\Pi_2)}{p'^2}\frac{\delta(p''-\Pi_1)}{p''^2}
\delta(\hat{p'}+\hat{\Pi_2})\delta(\hat{p''}-\hat{\Pi_1}),
\end{eqnarray}
then we write:
\begin{eqnarray} \label{eq:28}
&&\,^a\langle\textbf{p}\textbf{q}m_1m_2m_3\nu_1\nu_2\nu_3|tG_0P|U\rangle\nonumber
\\&&=2\sum_{m'_2m'_3\nu'_2\nu'_3}\int
d\textbf{q}''dp'dp''t_{m_2m_3m'_2m'_3}^{a,\nu_2\nu_3\nu'_2\nu'_3}(\textbf{p},p'\hat{\Pi_2})\nonumber
\\&&\times U_{m'_3m_1m'_2}^{\nu'_3\nu_1\nu'_2}
      (p''(-\hat{\Pi_1}),\textbf{q}'')\nonumber \\
&&\times\delta(p'-\Pi_2)\delta(p''-\Pi_1)\frac{1}{E-\frac{3q^2}{4m}-\frac{p'^2}{m}}.
\end{eqnarray}
in the above equation we used:
\begin{eqnarray} \label{eq:29}
&&t_{m_2m_3m'_2m'_3}^{a,\nu_2\nu_3\nu'_2\nu'_3}(\textbf{p},p'\hat{\Pi_2})\nonumber
\\&&=\,^a\langle\textbf{p}m_2m_3\nu_2\nu_3|t|p'\hat{\Pi_2}m'_2m'_3\nu'_2\nu'_3\rangle^a,
\end{eqnarray}
and
\begin{eqnarray} \label{eq:30}
&&U_{m_1m_2m_3}^{\nu_1\nu_2\nu_3}(p(-\hat{\Pi}_1),\textbf{q})\nonumber
\\&&=\,^a\langle
p(-\hat{\Pi}_1)\textbf{q}m_1m_2m_3\nu_1\nu_2\nu_3|U\rangle,
\end{eqnarray}
we have for the radial delta function:
\begin{eqnarray} \label{eq:31}
&&\delta(p'-\Pi_2)\delta(p''-\Pi_1)\nonumber
\\&&=\delta(p'-\sqrt{\frac{1}{4}q^2+q''^2+qq''x''}) \nonumber
\\&& \times\delta(p''-\sqrt{p'^2+\frac{3}{4}q^2-\frac{3}{4}q''^2}),
\end{eqnarray}
we can write by using the property of the delta function:
\begin{eqnarray} \label{eq:33}
\delta
(p'-\sqrt{\frac{1}{4}q^2+q''^2+qq''x''}\,)=\frac{2p'}{qq''}\delta(x''-x_0),
\end{eqnarray}
where
\begin{eqnarray} \label{eq:34}
x_0=\frac{p'^2-\frac{1}{4}q^2-q''^2}{qq''}.
\end{eqnarray}
$x_0$ is $\cos{\theta_{q''}}$ and it is expressed in this interval:
$-1\leq x_0\leq +1$.

  We use the Eqs. (\ref{eq:31}) and (\ref{eq:33}) in Eq. (\ref{eq:22}), therefore, the final form of Eq. (\ref{eq:26}) can be rewritten as
follows:
\begin{eqnarray} \label{eq:35.5}
&&^a\langle
\textbf{p}\textbf{q}m_1m_2m_3\nu_1\nu_2\nu_3|U\rangle\nonumber
\\&&=^a\langle
\textbf{p}\textbf{q}m_1m_2m_3\nu_1\nu_2\nu_3|(1+P)J|\psi\rangle\nonumber
\\&&+\frac{4}{q}\frac{1}{E-E_d-\frac{3q}{4m}}\sum_{m'_2m'_3\nu'_2\nu'_3}\int_0^{\infty}dp'\frac{p'}{E-\frac{3q'^2}{4m}-\frac{p'^2}{m}}
\nonumber \\ &&\times\int_{|\frac{q}{2}-p'|}^{\frac{q}{2}+p'}dq'' q''\int_{-1}^1 dx_{q''}\int_0^{2\pi}d\phi_{q''}\delta(x''-x_0)\nonumber \\
&&\times\hat{t}_{m_2m_3m'_2m'_3}^{a,\nu_2\nu_3\nu'_2\nu'_3}(\textbf{p},p'\hat{\Pi_2})U_{m'_3m_1m'_2}^{\nu'_3\nu_1\nu'_2}(p''(-\hat{\Pi_1}),\textbf{q}''),
\end{eqnarray}
where
\begin{eqnarray} \label{eq:38}
p''=\sqrt{p'^2-\frac{3q''^2}{4}+\frac{3q^2}{4}},
\end{eqnarray}
we have used the following relation to extract the singularity from
two-body t-matrix:
\begin{eqnarray} \label{eq:36}
\hat{t}(E)=\frac{t(E)}{E-E_d}.
\end{eqnarray}

\subsection{Current} \label{subsec:Faddeev-equations}
The current is consisted of single-nucleon and two-nucleon currents
\begin{eqnarray} \label{eq:55}
J&&=J^{I}+J^{II}, \nonumber \\
J^I&&=J^I(1)+J^I(2)+J^I(3),\nonumber\\
J^{II}&&=J^{II}(1,2)+J^{II}(2,3)+J^{II}(3,1).
\end{eqnarray}
According to the following symmetry relation one can only consider
 $J^I(1)$ and $J^{II}(2,3)$ in Eq. (\ref{eq:26}), i.e.
\begin{eqnarray} \label{eq:56}
&&^a\langle\textbf{p}\textbf{q}m_1m_2m_3\nu_1\nu_2\nu_3|(1+P)J|\Psi\rangle\nonumber
\\&&=3^a\langle\textbf{p}\textbf{q}m_1m_2m_3\nu_1\nu_2\nu_3|\nonumber \\&&(1+P)[J^I(1)+J^{II}(2,3)|\Psi\rangle,
\end{eqnarray}
the matrix elements of the single-nucleon current i.e.  $J^I(1)$ in
the free basis states is independent of the initial and final
two-body subsystem , and the initial and final spectators Jacobi
momentum vectors are related by the delta function
$\delta(\textbf{q}-\textbf{q}'-\frac{2}{3}\textbf{Q})$. So the
matrix element of the single-nucleon current takes the following
expression with the help of Eq. (\ref{eq:13}) and Eq. (\ref{eq:56})
and is evaluated as follows:
\begin{eqnarray} \label{eq:58}
&&^a\langle\textbf{p}\textbf{q}m_1m_2m_3\nu_1\nu_2\nu_3|(1+P)J^{I}|\Psi\rangle\nonumber
\\&&=3\sum_{m'_1\nu'_1}\bigg\{J_{m_1m'_1}^{\nu_1\nu'}(\textbf{q},\textbf{Q})\Psi_{m'_1m_2m_3}^{\nu'_1\nu_2\nu_3}
                                (\textbf{p},\textbf{q}-\frac{2}{3}\textbf{Q})\nonumber \\
&&+J_{m_2m'_1}^{\nu_2\nu'_1}(\textbf{p}-\frac{1}{2}\textbf{q},\textbf{Q})\Psi_{m'_1m_3m_1}^{\nu'_1\nu_3\nu_1}
(-\frac{1}{2}\textbf{p}-\frac{3}{4}\textbf{q},\textbf{p}-\frac{1}{2}\textbf{q}-\frac{2}{3}\textbf{Q})\nonumber \\
&&+J_{m_3m'_1}^{\nu_3\nu'_1}(-\textbf{p}-\frac{1}{2}\textbf{q},\textbf{Q})\nonumber
\\&&\times\Psi_{m'_1m_1m_2}^{\nu'_1\nu_1\nu_2}
(-\frac{1}{2}\textbf{p}+\frac{3}{4}\textbf{q},-\textbf{p}-\frac{1}{2}\textbf{q}-\frac{2}{3}\textbf{Q})\bigg\},
\end{eqnarray}

For two-body current we have:
\begin{eqnarray} \label{eq:60}
&&^a\langle\textbf{p}'\textbf{q}'m'_1m'_2m'_3\nu'_1\nu'_2\nu'_3|J(2,3)|\textbf{p}\textbf{q}m_1m_2m_3\nu_1\nu_2\nu_3\rangle^a\nonumber
\\&&=\delta(\textbf{q}'-\textbf{q}+\frac{1}{3}\textbf{Q})\delta_{m_1 m'_1}\delta_{\nu_1
\nu'_1}J_{m_2m_3m'_2m'_3}^{II,\nu_2\nu_3\nu'_2\nu'_3}(\textbf{p},\textbf{p}'),
\end{eqnarray}
so the two-body current with the help of Eq. (\ref{eq:56}) is:
\begin{eqnarray} \label{eq:61}
&&^a\langle\textbf{p}\textbf{q}m_1m_2m_3\nu_1\nu_2\nu_3|(1+P)J^{II}|\psi\rangle\nonumber
\\&&=3\sum_{m'_2m'_3\nu'_2\nu'_3}\int
d^3\textbf{p}'\nonumber
\\&&\times \bigg\{J_{m_2m_3m'_2m'_3}^{II,\nu_2\nu_3\nu'_2\nu'_3}(\textbf{p},\textbf{p}')\psi_{m_1m'_2m'_3}^{\nu_1\nu'_2\nu'_3}(\textbf{p}',\textbf{q}-\frac{1}{3}\textbf{Q})\nonumber
\\&& +J_{m_3m_1m'_2m'_3}^{II,\nu_3\nu_1\nu'_2\nu'_3}(-\frac{1}{2}\textbf{p}-\frac{3}{4}\textbf{q},\textbf{p}')\psi_{m_2m'_2m'_3}^{\nu_2\nu'_2\nu'_3}(\textbf{p}',\textbf{q}-\frac{1}{3}\textbf{Q})\nonumber
\\&&
+J_{m_1m_2m'_2m'_3}^{II,\nu_1\nu_2\nu'_2\nu'_3}(-\frac{1}{2}\textbf{p}+\frac{3}{4}\textbf{q},\textbf{p}')\psi_{m_3m'_2m'_3}^{\nu_3\nu'_2\nu'_3}(\textbf{p}',\textbf{q}-\frac{1}{3}\textbf{Q})\bigg\}.\nonumber
\\
\end{eqnarray}
For the meson exchange current we have taken into account the
momentum transfer of the two nucleons:
\begin{eqnarray} \label{eq:62}
\textbf{q}_2=\textbf{k}_2-\textbf{k}'_2=\frac{1}{2}\textbf{Q}+\textbf{p}-\textbf{p}',\nonumber
\\\textbf{q}_3=\textbf{k}_3-\textbf{k}'_3=\frac{1}{2}\textbf{Q}-\textbf{p}+\textbf{p}',
\end{eqnarray}

where $\textbf{k}_2$, $\textbf{k}'_2$, $\textbf{k}_3$,
$\textbf{k}'_3$ are the initial and final momenta of the nucleons 2
and 3.

 Single nucleon current in nonrelativistic limit as well as the two-body
meson exchange currents are introduced and evaluated in the free
basis states by suitable expressions in the Appendix
\ref{app:current}.
\section{Numerical methods and results} \label{sec:numerical}
In order to calculate the total cross sections by Eqs. (\ref{eq:1})
and (\ref{eq:1p}), we need to obtain the nuclear matrix elements
using Eqs. (\ref{eq:23}) and (\ref{eq:24}). In fact, we have to
calculate the auxiliary state $|U\rangle$ in the free basis stats
using Eq. (\ref{eq:35.5}). To solve this integral equation and also
Eqs. (\ref{eq:1}), (\ref{eq:1p}), (\ref{eq:23}) and (\ref{eq:24}),
we first need to choose the appropriate coordinate system.
 Because of the current properties, We have to choose the photon
momentum vector, $\textbf{Q}$, to be in $z$ direction, which is also
the spin-quantization axis. This selection does not impose any
restrictions because in all of the above equations $\textbf{Q}$ is a
constant vector. The $\textbf{p}$ and $\textbf{q}$ vectors like any
free vectors will be determined with three
 components. Finally, except for the energy of photon which is
 constant,
 we need six independent variables to specify U
 function in Eq. (\ref{eq:35.5}) in terms of $\textbf{p}$ and $\textbf{q}$ vectors.
\begin{eqnarray} \label{eq:65}
U(\textbf{p},\textbf{q})=U(p,q,\theta_p,\theta_q,\phi_p,\phi_q).
\end{eqnarray}

In appendix \ref{app:U function}, The relation between variables of
each term of Eq. (\ref{eq:35.5}) and six variables of U function is
demonstrated.

For the polar angles which vary from $0$ to $\pi$ the sign of the
sine is always positive and we can introduce $x_{\theta}=cos\theta$
as an independent variable. However, for the azimuthal angle, which
is in the interval $[0,2\pi]$, the sign of the sine can not be
specified as a function of $x_{\phi}=\cos\phi$ because in each
intervals of $0<\phi<\pi$ or $\pi<\phi<2\pi$ we have different signs
for sine. So it is difficult to specify the value of any function in
terms of $x_{\phi}=\cos\phi$.

In order to save computing time and computational memory we need to
define two functions for U in each interval.
\begin{eqnarray} \label{eq:66}
U(\phi)=\left\{\begin{array}{cc}
          U_1(\phi) & \mbox{if $0<\phi<\pi$} \\
          U_2(\phi) & \mbox{if $\pi<\phi<2\pi$}
        \end{array}\right.,
\end{eqnarray}
$U_1$ and $U_2$ are defined as:
\begin{eqnarray} \label{eq:67}
&&U_1(\phi)=F(cos\phi,\sqrt{1-cos^2\phi}),\\
&&U_2(\phi)=F(cos\phi,-\sqrt{1-cos^2\phi}),
\end{eqnarray}

 We write Eq. (\ref{eq:35.5}) in terms of the independent variables
 of the U function in an
 appropriate expression for numerical iteration. We consider
the U operator in a general form:
\begin{eqnarray} \label{eq:70}
|U\rangle=|U_0\rangle+K|U\rangle=|U_0\rangle+K|U_0\rangle+K^2|U_0\rangle+...,
\end{eqnarray}
where $|U_0\rangle$ is the current term and $K$ is an integral
kernel, then we apply the kernel $K$ to generate the finite Neumann
series up to $(i-1)th$ order in $K$. This Neumann series can be
summed up using Pade approximation to get $U_{Pade}^{(i-1)}$
\cite{H. Liu}. We get $U_{Pade}^{(i)}$ from one more iteration. The
definition of the distance between $U_{Pade}^{(i-1)}$ and
$U_{Pade}^{(i)}$ is:
\begin{eqnarray} \label{eq:71}
\Delta^i=\frac{\sum|U_{Pade}^{(i)}-U_{Pade}^{(i-1)}||U_{Pade}^{(i)}|}{\sum{|U_{Pade}^{(i)}|}^2}.
\end{eqnarray}
where, the summation runs over all six-dimensional grid points. We
continue the iteration to reach $\Delta^n<\epsilon$. $\epsilon$ is a
small number determined by the desired accuracy. In our work $n=10$
and $\epsilon=10^{-2}$.

According to Eqs. (\ref{eq:35.5}) and (\ref{eq:13}) we need the
two-body t-matrix and triton wave function in the free basis states.
The extraction of singularity from two-body t-matrix in deuteron
binding energy (appendix \ref{app:Deuteron}) shows that we also need
the deuteron wave function in the free basis states. For calculation
of the two-body t-matrix, deuteron and triton wave functions we
follow the 3D approach introduced in appendix \ref{app:tmatrix
deuteron and triton} and we recalculate them in the free basis
states using AV18 potential \cite{AV18}. In the calculation of the
Eq. (\ref{eq:35.5}) we interpolate the calculated data of the
t-matrix, deuteron and triton wave functions using Cubic-Hermit
spline method \cite{spline}. In the numerical treatment the momenta
and angles variables should be transformed to certain discrete
values. For the AV18 potential we use the Gaussian quadrature grid
points with the hyperbolic mapping for the lower momentum and linear
mapping for the higher momentum.  The numerical Fourier-Bessel
transformation of the potential encounters difficulties in handling
at the very high momentum, so it is necessary to use a cut off in
the integration interval at $150 fm^{-1}$.

The meson exchange currents (MECs) were restricted to $\pi$- like
and $\rho$- like exchanges. These MECs are derived from AV18 Based
on the Riska's recipe \cite{riska}.

We compare the 3D Faddeev calculation, including the explicit MECs
and the PW representation of Faddeev calculation, with Siegert
theorem \cite{seigert}. We have shown that the 3D approach with the
continuous angle variables instead of the discrete angular momentum
quantum numbers in evaluation of the nuclear matrix elements for
$\gamma\,^3 H\rightarrow Nd$ and $\gamma\,^3 H\rightarrow 3N$ leads
to less complicated expressions but with higher dimensionality of
integral equations in comparison with the PW representation.

\begin{figure*}
\resizebox{0.75\textwidth}{!}{%
\includegraphics{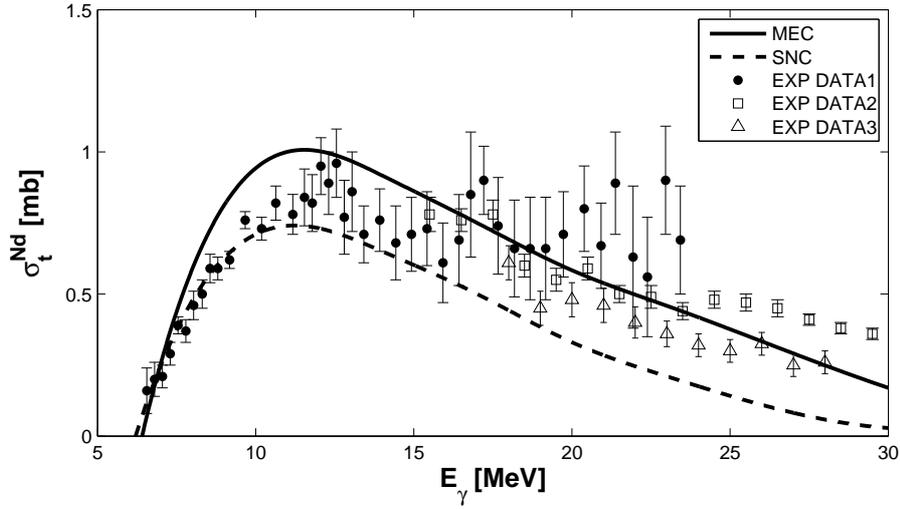}
} \caption{Comparison of the 3D calculations for the total cross
section of the two-body (Nd) photodisintegration of Triton using the
single nucleon current (dashed line) and the two-body current (solid
line). The experimental values are taken from Ref.\cite{Faul_PhRC24}
as EXP. DATA1, \cite{Skopik_PhRC24} as EXP. DATA2 and
\cite{Kosiek_PhL21} as EXP. DATA3.}
\label{fig1.Single Nucleon Nd}       
\end{figure*}

\begin{figure*}
\resizebox{0.75\textwidth}{!}{%
\includegraphics{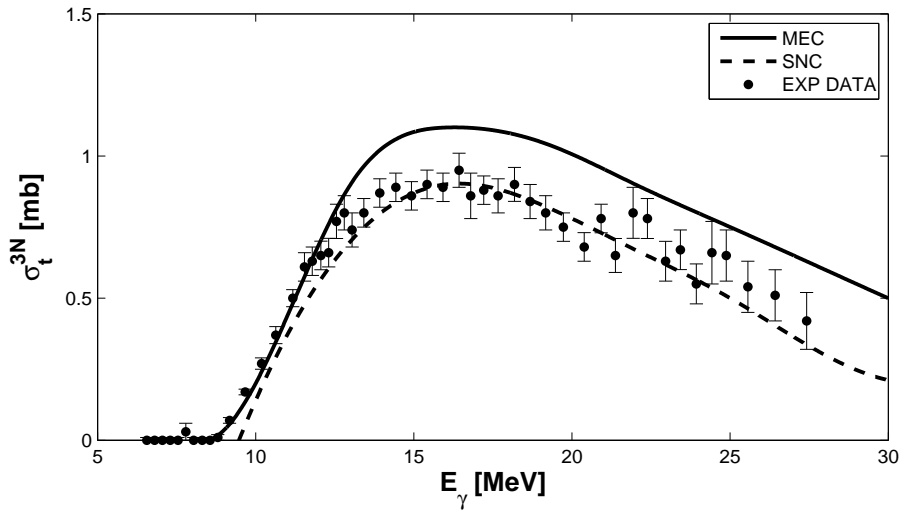}
} \caption{Comparison of the 3D calculations for the total cross
section of the three-body (3N) photodisintegration of Triton using
the single nucleon current (dashed line) the two-body current (solid
line). The experimental values are taken from Ref.
\cite{Faul_PhRC24}.}
\label{fig2.Single Nucleon 3N}       
\end{figure*}

\begin{figure*}
\vspace*{5cm}       
\resizebox{0.75\textwidth}{!}{%
\includegraphics{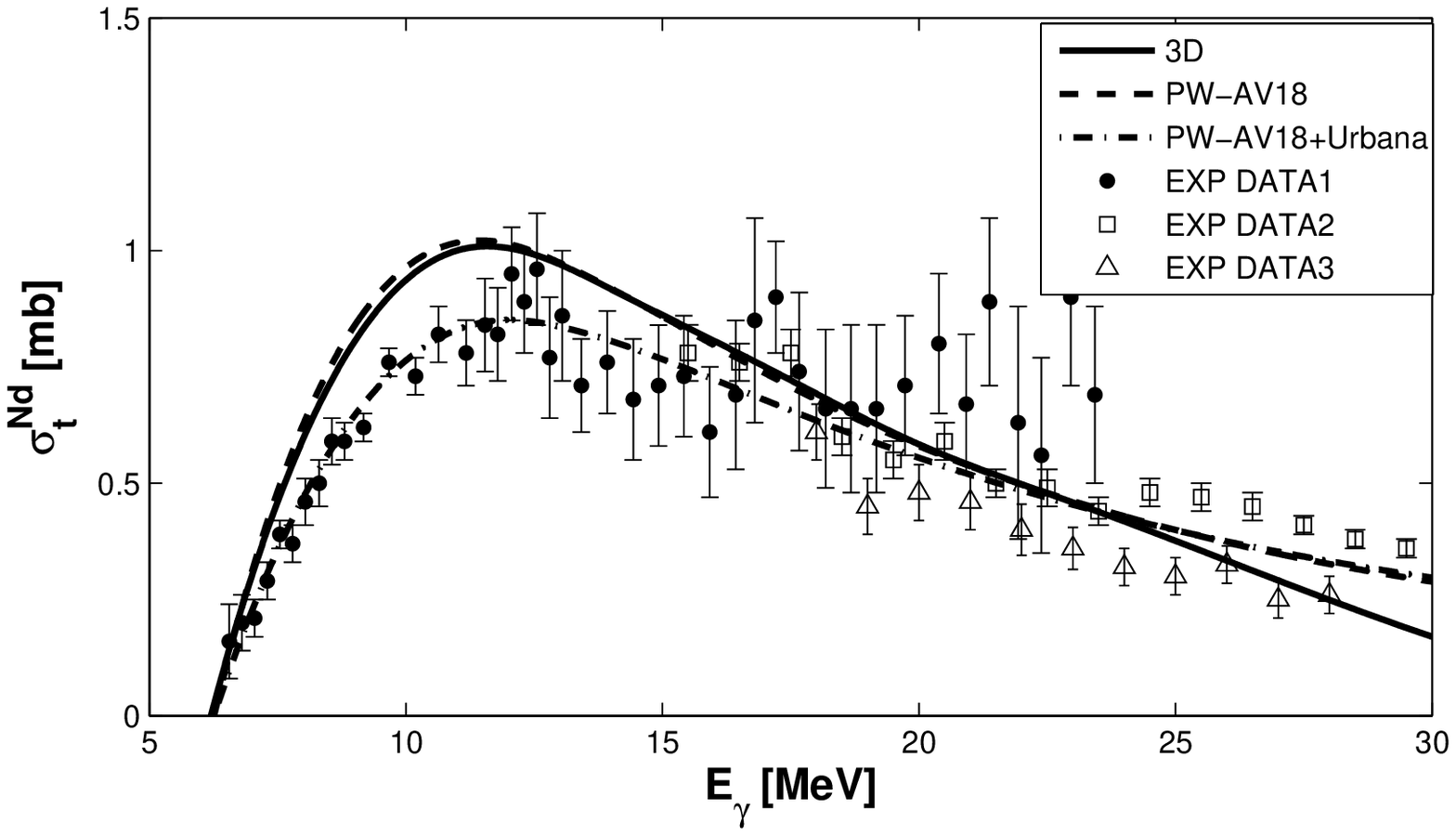}
} \caption{Total cross section for Nd photodisintegration of Triton.
The solid line is the 3D calculation with AV18 potential.The dashed
line represents the partial wave calculation using AV18 potential
and dot-dashed line represents the partial wave result using AV18
potential and 3NF (Urbana IX). The experimental values are the same
as Fig. (\ref{fig1.Single Nucleon Nd}).}
\label{fig3.Nd cross section}       
\end{figure*}

\begin{figure*}
\resizebox{0.75\textwidth}{!}{%
\includegraphics{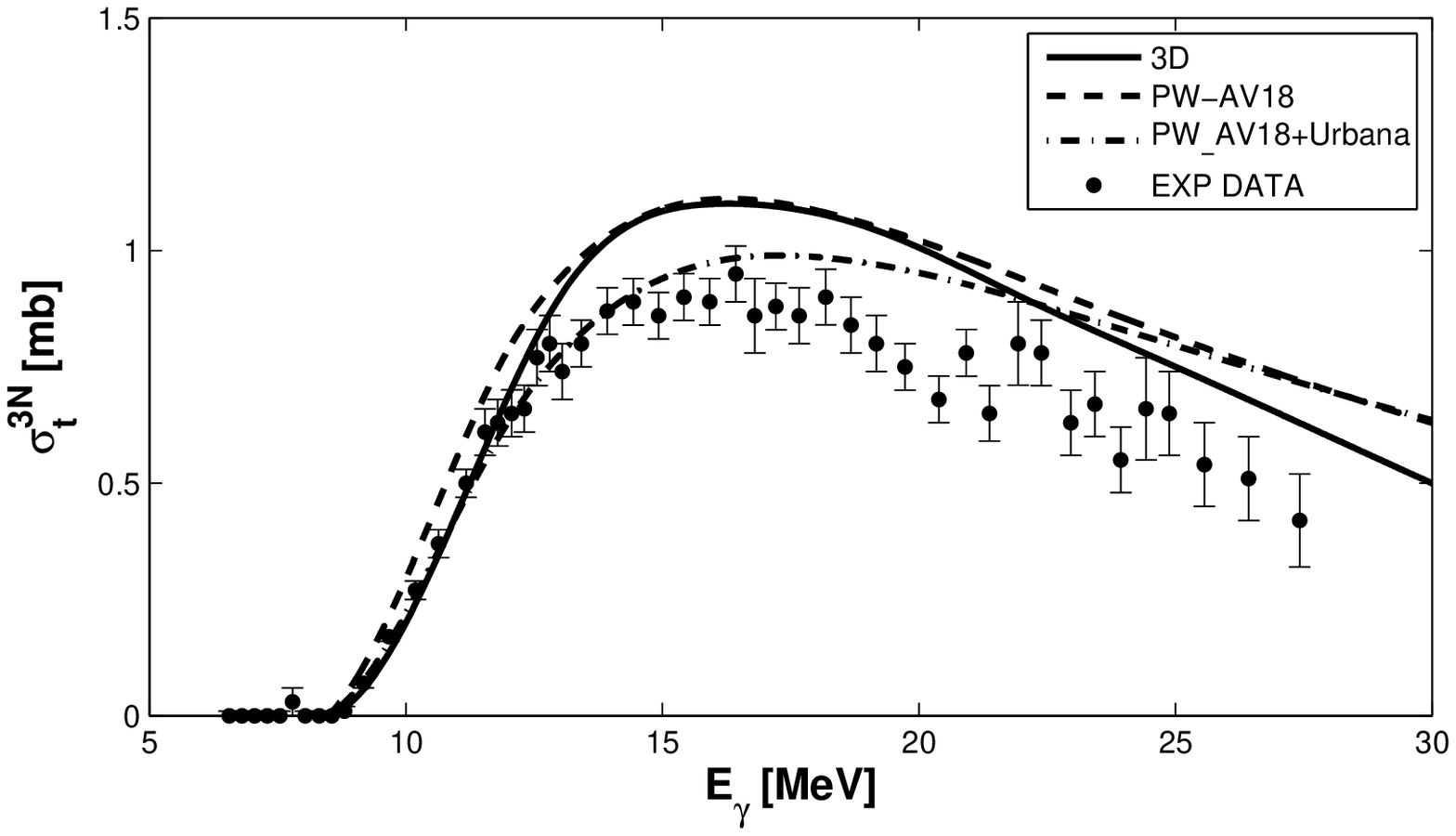}
}\caption{Total cross section for 3N photodisintegration of Triton.
The solid line is the 3D calculation with AV18 potential.The dashed
line represents the partial wave calculation using the AV18
potential and dot-dashed line represents the partial wave result
using AV18 potential and 3NF (Urbana IX). The experimental values
are the same as Fig. (\ref{fig2.Single Nucleon 3N}).}
\label{fig4.3N cross section}       
\end{figure*}

 Calculation of the one-body current and the two-body current for
$\sigma_t^{Nd}$ and $\sigma_t^{3N}$ are compared together, with
experimental data in Figs. (\ref{fig1.Single Nucleon Nd}) and
(\ref{fig2.Single Nucleon 3N}). The comparison re-confirms the
enhancement of the three-nucleon photodisintegration cross section
in the peak region and at the higher energies. The addition of the
three-nucleon force can significantly lower the peak in
$\sigma_t^{Nd}$ and $\sigma_t^{3N}$ and as a result gives better
agreement with the available data. We have displayed in Figs.
(\ref{fig3.Nd cross section}) and (\ref{fig4.3N cross section}) that
the Siegert and MEC predictions are too close in the lower energies
of the photodisintegration of $^3$H to $Nd$ and $3N$. However, the
3D calculation uses the single nucleon current with explicit use of
$\pi$- and $\rho$- like mesons. We found discrepancies between the
two predictions in the higher energies of the photodisintegration.
We expect that in the higher energies of the photon, the use of
meson exchange in 3D approach produces more sensible results.

In comparison with the experimental data (Ref \cite{Faul_PhRC24}),
Fig. (\ref{fig3.Nd cross section}), the 3D total cross section of
$\gamma\,^3 H\rightarrow Nd$ shows less agreement at low energies
($E_{\gamma}<20 MeV$). The overestimation shows the need for a
three-nucleon force (3NF) effect. The results of Golak et. al
\cite{benchmark} indicate the improvement by adding 3NF. The
contribution of 3NF in the 3D calculations can be implemented by
modifying the $|U\rangle$ as an auxiliary state with the term
appropriate for the 3NF adding to the two-body forces.

The data of (Ref \cite{Faul_PhRC24}) at $E_{\gamma}=20-30 MeV$ due
to the insufficient precision can not be compared with the
theoretical calculations and no concrete conclusions can be reached.

 The Skopic et. al data and Kosiake et al data \cite{Skopik_PhRC24,Kosiek_PhL21}
 although nearly agree with the 3D calculations in ($E_{\gamma}=15-28 MeV$),
however more experimental data is needed to reach better conclusion.

In the case of $\gamma\,^3 H\rightarrow3N$ the overestimation of the
calculated total cross section in the low and medium energies in
Fig. (4) is also predicted to be due to the absence of 3NF. At
higher energies more experimental data is needed to overcome the
discrepancies with the present day theories.

\section{Summary and outlook} \label{sec:summary}

In this paper we have formulated the Faddeev integral equations for
 calculating the two- and three-body photodisintegration cross sections of the triton in a 3D approach.
 To this aim we have used the free basis states
 which contain Jacobi momentum vectors as well as individual
 spin and isospin of the nucleons. So we avoid to decompose
 the angular dependent in terms of the angular momentum quantum
 numbers,
  traditionally used to solve these kind of equations, i.e. partial wave approach. The
 final integral equations are less complicated than the
 similar partial wave integral equations and are unique in number of the equations in all energies. We
 have also explained  how to overcome the moving
 singularity in Eq. (\ref{eq:26}) by the separation of the radial and angle parts of the Dirac delta function.

 Using Eqs. (\ref{eq:1}) and (\ref{eq:1p}) we have calculated the total cross section for
 3N and Nd photodisintegration of the triton. Benchmarks for the three-nucleon total photodisintegration cross
 sections
 are presented in Figs. (\ref{fig3.Nd cross section}) and (\ref{fig4.3N
cross section}).

Although the classical approximation of photodisintegration cross
section (predicted by Golak et. al \cite{benchmark}) and the 3D
calculation with MECs are nearly in agreement for both $\gamma\,^3
H\rightarrow Nd$ and $\gamma\,^3 H\rightarrow 3N$ at low energies,
the significance of the 3D based calculation with MECs can be tested
further with the inclusion of the 3NF calculation.

Adding the three-body current as well as the three-body forces  in
our calculations are the other major future works to be done. We
have calculated the two-body t-matrices using chiral potential in
the 3D approach \cite{Bayegan_PhRC79}. Therefore the calculation of
three-body photodisintegration by this potential  using the chiral
currents is another area for consideration. The similar calculation
for the radiative capture is also under preparation.

\section*{Acknowledgments}
We would like to thank J. Golak and R. Skibinski for providing us
the results of their calculations. This work was supported by the
research council of the university of Tehran.

\appendix

\section{matrix elements of current} \label{app:current}

Single nucleon current in nonrelativistic limit consists of
convection and spin current terms:
\begin{eqnarray} \label{eq:59}
J(1)=G_E(Q)\frac{\textbf{k}_1+\textbf{k}'_1}{2m_N}+\frac{i}{2m_N}G_M(Q)\vec{\sigma}\times(\textbf{k}'_1-\textbf{k}_1).\;\;\;\;
\end{eqnarray}
Where $G_E$ and $G_M$ are electric and magnetic form factors of the
nucleon, respectively. $\textbf{k}_1$ and $\textbf{k}'_1$ are the
initial and final momentum of nucleon 1.

We used the matrix element of the one-body currents in the tensor
component representation. Considering this equality:
\begin{eqnarray} \label{eq:b1}
\textbf{k}_1+\textbf{k}'_1&&=2\textbf{q}-\textbf{Q}+\frac{2}{3}(\textbf{k}_1+\textbf{k}_2+\textbf{k}_3)\nonumber
\\&&=2\textbf{q}-\textbf{Q}+\frac{2}{3}\textbf{K},
\end{eqnarray}
we can rewrite the single nucleon current in the representation of
the tensor components as follows:
\begin{eqnarray} \label{eq:b2}
J_{\pm}^{convec}=\frac{q_{\pm1}}{m_N}(G_E^p(Q)\Pi^p+G_E^n(Q)\Pi^n),\\
J_{\pm}^{Spin}=-\frac{\sqrt{2}Q}{2m_N}S_{\pm}(G_M^p(Q)\Pi^p+G_M^n(Q)\Pi^n),
\end{eqnarray}
where
\begin{eqnarray} \label{eq:b3}
\Pi^p=|p\rangle\langle p|,~~~~~~~~~~~~\Pi^n=|n\rangle\langle n|,\\
S_+=|+\rangle\langle+|~~~~~~~~~~~~~~~S_-=|-\rangle\langle-|,
\end{eqnarray}
and finally:
\begin{eqnarray} \label{eq:b4}
J_{m_1m'_1}^{+\,,\nu_1\nu'_1}(\textbf{q},\textbf{Q})=\left\{\begin{array}{ccc}
                               0 & \mbox{for $\nu\neq\nu'$} \\
                               0 & \mbox{for $m'=m+1$} \\
                               -\frac{qsin\theta_q}{\sqrt{2}m_N}e^{i\phi_q}G_E^{\nu}(Q) & \mbox{for $m'=m$} \\
                               -\frac{\sqrt{2}}{2m_N}G_M^{\nu}(Q) & \mbox{for
                               $m'=m-1$}
                             \end{array}\right.,\;\;\;\;
\\J_{m_1m'_1}^{-\,,\nu_1\nu'_1}(\textbf{q},\textbf{Q})=\left\{\begin{array}{ccc}
                               0 & \mbox{for $\nu\neq\nu'$} \\
                               -\frac{\sqrt{2}}{2m_N}G_M^{\nu}(Q) & \mbox{for $m'=m+1$} \\
                               \frac{qsin\theta_q}{\sqrt{2}m_N}e^{-i\phi_q}G_E^{\nu}(Q) & \mbox{for $m'=m$} \\
                                0 & \mbox{for $m'=m-1$}
                             \end{array}\right..\;\;\;\;
\end{eqnarray}

The following $\pi-$ and $\rho-$ meson exchange currents are
introduced in \cite{golak} as follows:
\begin{eqnarray} \label{eq:63}
&&\vec{j}_{\pi}(\textbf{q}_2,\textbf{q}_3)=i\bigg(G_E^p(Q)-G_E^n(Q)\bigg)(\vec{\tau_2}\times\vec{\tau_3})_3
\nonumber
\\ &&\times\bigg[\vec{\sigma_3}\vec{\sigma_2}\cdot\textbf{q}_2v_{\pi}(q_2)-\vec{\sigma_2}\vec{\sigma_3}\cdot\textbf{q}_3v_{\pi}(q_3)\nonumber
\\&&+\frac{\textbf{q}_2-\textbf{q}_3}{q_2^2-q_3^2}\bigg(v_{\pi}(q_3)-v_{\pi}(q_2)\bigg)
\vec{\sigma_2}\cdot\textbf{q}_2\vec{\sigma_3}\cdot\textbf{q}_3\bigg],
\end{eqnarray}
and
\begin{eqnarray} \label{eq:64}
&&\vec{j}_{\rho}(\textbf{q}_2,\textbf{q}_3)=i\bigg(G_E^p(Q)-G_E^n(Q)\bigg)(\vec{\tau_2}\times\vec{\tau_3})_3\nonumber
\\ &&\bigg[\frac{\textbf{q}_2-\textbf{q}_3}{q_2^2-q_3^2}\bigg(v_{\rho}^S(q_3)-v_{\rho}^S(q_2)\bigg)\nonumber
\\&&-\bigg(v_{\rho}(q_3)\vec{\sigma_2}\times(\vec{\sigma_3}\times\textbf{q}_3)-v_{\rho}(q_2)\vec{\sigma_3}\times(\vec{\sigma_2}\times\textbf{q}_2)\bigg)\nonumber
\\&&-\frac{v_{\rho}(q_3)-v_{\rho}(q_2)}{q_2^2-q_3^2}\bigg((\vec{\sigma_2}\times\textbf{q}_2)\cdot(\vec{\sigma_3}\times\textbf{q}_3)(\textbf{q}_2-\textbf{q}_3)\nonumber
\\&&-\vec{\sigma_3}\cdot(\textbf{q}_2\times\textbf{q}_3)(\vec{\sigma_2}\times\textbf{q}_2)\nonumber \\&&-\vec{\sigma_2}\cdot(\textbf{q}_2\times\textbf{q}_3)(\vec{\sigma_3}\times\textbf{q}_3)\bigg)\bigg].
\end{eqnarray}
The functions $v_{\pi}(q)$, $v_{\rho}^S(q)$ and $v_{\rho}(q)$ can be
extracted from the phenomenological AV18 two-nucleon interaction
\cite{golak}.

The matrix elements of the two-body current are also written in the
representation of the tensor component. For the pion-exchange we can
write:
\begin{eqnarray} \label{eq:b5}
&&\langle\textbf{p}'\textbf{q}'m'_1m'_2m'_3\nu'_1\nu'_2\nu'_3|J_{\pm}^{\pi}|\textbf{p}\textbf{q}m_1m_2m_3\nu_1\nu_2\nu_3\rangle\nonumber
\\
&&=\delta_{m_1m'_1}\delta_{\nu_1\nu'_1}\delta(\textbf{q}'-\textbf{q}+\frac{1}{3}\textbf{Q})\bigg(G_E^p(Q)-G_E^n(Q)\bigg)\nonumber
\\
&&(\delta_{\nu'_2,\nu_2-1}\delta_{\nu'_3,\nu_3+1}-\delta_{\nu'_2,\nu_2+1}\delta_{\nu'_3,\nu_3-1})\nonumber\\
&&\times\bigg\{(\mp2\sqrt{2}\delta_{m'_3,m_3\pm1}\bigg(\sum_{m''}m''D_{m'_2m''}(\hat{q}_2)D_{m''m_2}^+(\hat{q}_2)\bigg)v_{\pi}(q_2)\nonumber
\\&&\pm2\sqrt{2}\delta_{m'_2,m_2\pm1}\bigg(\sum_{m''}m''D_{m'_3m''}(\hat{q}_3)D_{m''m_3}^+(\hat{q}_3)\bigg)v_{\pi}(q_3)\nonumber
\\&&+\frac{(\textbf{q}_2-\textbf{q}_3)_{\pm1}}{q_2^2-q_3^3}(v_{\pi}(q_3)-v_{\pi}(q_2))\nonumber
\\
&&\times4\bigg(\sum_{m''}m''D_{m'_2m''}(\hat{q}_2)D_{m''m_2}^+(\hat{q}_2)\bigg)\nonumber
\\&&\bigg(\sum_{m''}m''D_{m'_3m''}(\hat{q}_3)D_{m''m_3}^+(\hat{q}_3)\bigg)\bigg\}.
\end{eqnarray}

For the matrix elements of $\rho-$exchange current in the tensor
components form we can write:
\begin{eqnarray} \label{eq:b6}
&&\langle\textbf{p}'\textbf{q}'m'_1m'_2m'_3\nu'_1\nu'_2\nu'_3|J_{\pm}^{\rho}|\textbf{p}\textbf{q}m_1m_2m_3\nu_1\nu_2\nu_3\rangle\nonumber
\\&&=\delta_{m_1m'_1}\delta_{\nu_1\nu'_1}\delta(\textbf{q}'-\textbf{q}+\frac{1}{3}\textbf{Q})\bigg(G_E^p(Q)-G_E^n(Q)\bigg)\nonumber
\\
&&(\delta_{\nu'_2,\nu_2-1}\delta_{\nu'_3,\nu_3+1}-\delta_{\nu'_2,\nu_2+1}\delta_{\nu'_3,\nu_3-1})\nonumber
\\&& \times \bigg\{\frac{(\textbf{q}_2-\textbf{q}_3)_{\pm1}}{q_2^2-q_3^3}\bigg(v_{\rho}^s(q_3)-v_{\rho}^s(q_2)\bigg)\nonumber
\\&&-(\mp2\sqrt{2}\delta_{m'_3,m_3\pm1}\bigg(\sum_{m''}m''D_{m'_2m''}(\hat{q}_3)D_{m''m_2}^+(\hat{q}_3)\bigg)v_{\rho}(q_3)\nonumber
\\&&-(\delta_{m'_2,m_2+1}\delta_{m'_3,m_3-1}+\delta_{m'_2,m_2-1}\delta_{m'_3,m_3+1}\nonumber
\\&&+2m_2m_3\delta_{m_2m'_2}\delta_{m_3m'_3})(2q_{3,\pm1}v_{\rho}(q_3)-2q_{2,\pm1}v_{\rho}(q_2))\nonumber
\\&&\pm2\sqrt{2}\delta_{m'_2,m_2\pm1}\bigg(\sum_{m''}m''D_{m'_3m''}(\hat{q}_2)D_{m''m_3}^+(\hat{q}_2)\bigg)v_{\rho}(q_2)\nonumber
\\
&&-\frac{v_{\rho}(q_3)-v_{\rho}(q_2)}{q_3^2-q_2^2}\bigg[(\textbf{q}_2\cdot\textbf{q}_3)
\bigg(2(\delta_{m'_2,m_2+1}\delta_{m'_3,m_3-1}\nonumber
\\&&+\delta_{m'_2,m_2-1}\delta_{m'_3,m_3+1}+2m_2m_3\delta_{m'_2,m_2}\delta_{m'_3,m_3})\nonumber
\\&&\bigg(\sum_{m''}2m''D_{m'_2m''}(\hat{q}_2)D_{m''m_2}^+(\hat{q}_2)\bigg)\nonumber \\&&\bigg(\sum_{m''}2m''D_{m'_3m''}(\hat{q}_3)D_{m''m_3}^+(\hat{q}_3)\bigg)\bigg)\times(\textbf{q}_2-\textbf{q}_3)_{\pm1}\nonumber
\\&&-q_{3,\pm1}\bigg(2\sum_{m''}m''D_{m'_2m''}(\hat{q}_2)D_{m''m_2}^+(\hat{q}_2)\bigg)\nonumber
\\&&\bigg(2\sum_{m''}m''D_{m'_3m''}(\hat{q}_2)D_{m''m_3}^+(\hat{q}_2)\bigg)\nonumber
\\&&\pm\sqrt{2}(\textbf{q}_2\cdot\textbf{q}_3)\delta_{m'_2,m_2\pm1}\bigg(2\sum_{m''}m''D_{m'_3m''}(\hat{q}_2)D_{m''m_3}^+(\hat{q}_2)\bigg)\nonumber
\\
&&+q_{2,\pm1}\bigg(2\sum_{m''}m''D_{m'_2m''}(\hat{q}_3)D_{m''m_2}^+(\hat{q}_3)\bigg)\nonumber
\\&&\bigg(2\sum_{m''}m''D_{m'_3m''}(\hat{q}_3)D_{m''m_3}^+(\hat{q}_3)\bigg)\nonumber
\\&&\pm\sqrt{2}(\textbf{q}_2\cdot\textbf{q}_3)\delta_{m'_3,m_3\pm1}\bigg(2\sum_{m''}m''D_{m'_2m''}(\hat{q}_2)D_{m''m_2}^+(\hat{q}_3)\bigg)\bigg]\nonumber
\\&&+(\textbf{q}_2\times\textbf{q}_3)_{\pm1}
\bigg[\bigg((\delta_{m'_3,m_3+1}-\delta_{m'_3,m_3-1})\delta_{m'_2,m_2}\nonumber
\\&&-\delta_{m'_3,m_3}(\delta_{m'_2,m_2+1}-\delta_{m'_2,m_2-1})\bigg)\frac{q_{2x}-q_{3x}}{i}\nonumber
\\&&+\bigg((\delta_{m'_2,m_2+1}+\delta_{m'_2,m_2-1})\delta_{m'_3,m_3}\nonumber \\&&-\delta_{m'_2,m_2}(\delta_{m'_3,m_3+1}+\delta_{m'_3,m_3-1})\bigg)(q_{2y}-q_{3y})\nonumber
\\&&+(\delta_{m'_2,m_2+1}\delta_{m'_3,m_3-1}-\delta_{m'_2,m_2-1}\delta_{m'_3,m_3+1})\nonumber \\&&\times\frac{q_{2z}-q_{3z}}{i}\bigg]\bigg\}.
\end{eqnarray}

In the above equations $\textbf{q}_2$ and $\textbf{q}_3$ are
introduced in Eq. (\ref{eq:62}) and $D_{m'm}(\hat{q})$ is rotation
matrix element for $j=\frac{1}{2}$, $\alpha=\phi_q$,
$\beta=\theta_q$ and $\gamma=0$:

\begin{eqnarray} \label{eq:b7}
D_{m'm}^j(\alpha\beta\gamma)&&=\langle
jm'|R(\alpha\beta\gamma)|jm\rangle
\\&&=e^{-im'\alpha}d_{m'm}^j(\beta)e^{-im\gamma}\nonumber.
\end{eqnarray}

\section{U function equation in details} \label{app:U function}

The U function Eq. (\ref{eq:35.5}) consists of single nucleon
current ($SNC$), two-nucleon current ($TBC$) and a complicated part
denoting by($I$).

\begin{eqnarray} \label{eq:c2}
&&U_{\pm,m_1m_2m_3}^{\nu_1\nu_2\nu_3}(p,q,x_p,x_q,x_{\phi_p},x_{\phi_q})\nonumber
\\&&=SNC+TBC+I,
\end{eqnarray}

For the $SNC$ part we can incorporate the one-nucleon current as
follows:
\begin{eqnarray} \label{eq:c3}
&&SNC\nonumber
\\&&=3\sum_{m'_1\nu'_1}\bigg[J_{\pm,m_1m'_1}^{I,\nu_1\nu'_1}(q,x_q,x_{\phi_q})\psi_{m'_1m_2m_3}^{\nu'_1\nu_2\nu_3}(p,q',x_p,x_q,x_{\phi_p},x_{\phi_q})\nonumber
\\&&+J_{\pm,m_2m'_1}^{I,\nu_2\nu'_1}(q_2,x_{2q},x_{2\phi_q})\psi_{m'_1m_3m_1}^{\nu'_1\nu_3\nu_1}(p_2,q'_2,x_{2p},x'_{2q},x_{2\phi_p},x_{2\phi_q})\nonumber
\\&&+J_{\pm,m_3m'_1}^{I,\nu_3\nu'_1}(q_3,x_{3q},x_{3\phi_q})\nonumber \\&&\times\psi_{m'_1m_1m_2}^{\nu'_1\nu_1\nu_2}(p_3,q'_3,x_{3p},x'_{3q},x_{3\phi_p},x_{3\phi_q})\bigg],
\end{eqnarray}

In term of the $p,q,x_p,x_q,x_{\phi_p},x_{\phi_q}$ we can write all
the variable as follows:
\begin{eqnarray} \label{eq:c4}
q'=|\textbf{q}-\frac{2}{3}\textbf{Q}|=(q^2+\frac{4}{9}Q^2-\frac{4}{3}qQx_q)^{\frac{1}{2}},
\end{eqnarray}
\begin{eqnarray} \label{eq:c5}
p_2=|-\frac{1}{2}\textbf{p}-\frac{3}{4}\textbf{q}|=(\frac{1}{4}p^2+\frac{9}{16}q^2+\frac{3}{4}pqcos\gamma)^{\frac{1}{2}},
\end{eqnarray}
\begin{eqnarray} \label{eq:c6}
q_2=|\textbf{p}-\frac{1}{2}\textbf{q}|=(p^2+\frac{1}{4}q^2+pqcos\gamma)^{\frac{1}{2}},
\end{eqnarray}
\begin{eqnarray} \label{eq:c7}
\cos\gamma&&=x_px_q+\sqrt{1-x_p^2}\sqrt{1-x_q^2}\nonumber
\\&&\bigg(x_{\phi_p}x_{\phi_q}+\sqrt{1-x_{\phi_p}^2}\sqrt{1-x_{\phi_q}^2}\bigg),
\end{eqnarray}
\begin{eqnarray} \label{eq:c8}
x_{2q}=\frac{px_p-\frac{1}{2}qx_q}{q_2},
\end{eqnarray}
\begin{eqnarray} \label{eq:c9}
x_{2p}=\frac{-\frac{1}{2}px_p-\frac{3}{4}qx_q}{p_2},
\end{eqnarray}
\begin{eqnarray} \label{eq:c10}
q'_2&&=|\textbf{p}-\frac{1}{2}\textbf{q}-\frac{2}{3}\textbf{Q}|\nonumber
\\
&&=(p^2+\frac{1}{4}q^2+\frac{4}{9}Q^2-pq\cos\gamma\nonumber
\\&&-\frac{4}{3}pQx_p+\frac{2}{3}qQx_q)^{\frac{1}{2}},
\end{eqnarray}
\begin{eqnarray} \label{eq:c11}
x'_{2q}=\frac{px_p-\frac{1}{2}qx_q-\frac{2}{3}Q}{q'_2},
\end{eqnarray}
\begin{eqnarray} \label{eq:c12}
x_{2\phi_p}&&=\cos\phi_{2p}\nonumber
\\&&=\frac{-\frac{1}{2}p\sqrt{1-x_p^2}x_{\phi_p}-\frac{3}{4}q\sqrt{1-x_q^2}x_{\phi_q}}{p_2\sqrt{1-x_{2p}^2}},
\end{eqnarray}
\begin{eqnarray} \label{eq:c13}
&&sin\phi_{2p}\nonumber
\\&&=\bigg(-\frac{1}{2}p\sqrt{1-x_p^2}\sqrt{1-x_{\phi_p}^2}-\frac{3}{4}q\sqrt{1-x_q^2}\sqrt{1-x_{\phi_q}^2}\bigg)\nonumber
\\&&/\bigg(p_2\sqrt{1-x_{2p}^2}\bigg),
\end{eqnarray}
\begin{eqnarray} \label{eq:c12}
x_{2\phi_q}=\cos\phi_{2q}=\frac{p\sqrt{1-x_p^2}x_{\phi_p}-\frac{1}{2}q\sqrt{1-x_q^2}x_{\phi_q}}{q_2\sqrt{1-x_{2q}^2}},
\end{eqnarray}
\begin{eqnarray} \label{eq:c13}
\sin\phi_{2q}=&&\bigg(p\sqrt{1-x_p^2}\sqrt{1-x_{\phi_p}^2}\nonumber
\\&&-\frac{1}{2}q\sqrt{1-x_q^2}\sqrt{1-x_{\phi_q}^2}\bigg)\nonumber
\\&&/{q_2\sqrt{1-x_{2q}^2}},
\end{eqnarray}
\begin{eqnarray} \label{eq:c14}
p_3=|-\frac{1}{2}\textbf{p}+\frac{3}{4}\textbf{q}|=(\frac{1}{4}p^2+\frac{9}{16}q^2-\frac{3}{4}pq\cos\gamma)^{\frac{1}{2}},
\end{eqnarray}
\begin{eqnarray} \label{eq:c15}
q_3=|-\textbf{p}-\frac{1}{2}\textbf{q}|=(p^2+\frac{1}{4}q^2+pqcos\gamma)^{\frac{1}{2}},
\end{eqnarray}
\begin{eqnarray} \label{eq:c16}
x_{3q}=\frac{-px_p-\frac{1}{2}qx_q}{q_3},
\end{eqnarray}
\begin{eqnarray} \label{eq:c17}
x_{3p}=\frac{-\frac{1}{2}px_p+\frac{3}{4}qx_q}{p_2},
\end{eqnarray}
\begin{eqnarray} \label{eq:c18}
q'_3&&=|-\textbf{p}-\frac{1}{2}\textbf{q}-\frac{2}{3}\textbf{Q}|\nonumber
\\
&&=(p^2+\frac{1}{4}q^2+\frac{4}{9}Q^2+pqcos\gamma\nonumber
\\&&+\frac{4}{3}pQx_p+\frac{2}{3}qQx_q)^{\frac{1}{2}},
\end{eqnarray}
\begin{eqnarray} \label{eq:c19}
&&x_{3\phi_p}=\cos\phi_{3p}\nonumber
\\&&=\bigg(-\frac{1}{2}p\sqrt{1-x_p^2}x_{\phi_p}+\frac{3}{4}q\sqrt{1-x_q^2}x_{\phi_q}\bigg)\nonumber
\\&&/\bigg(p_2\sqrt{1-x_{3p}^2}\bigg),
\end{eqnarray}
\begin{eqnarray} \label{eq:c20}
&&\sin\phi_{3p}\nonumber
\\&&=\bigg(-\frac{1}{2}p\sqrt{1-x_p^2}\sqrt{1-x_{\phi_p^2}}+\frac{3}{4}q\sqrt{1-x_q^2}\sqrt{1-x_{\phi_q}^2}\bigg)\nonumber
\\&&/{p_2\sqrt{1-x_{3p}^2}}.
\end{eqnarray}
\begin{eqnarray} \label{eq:c19}
&&x_{3\phi_q}=\cos\phi_{3q}\nonumber
\\&&=\bigg(-p\sqrt{1-x_p^2}x_{\phi_p}-\frac{1}{2}q\sqrt{1-x_q^2}x_{\phi_q}\bigg)\nonumber
\\&&/\bigg(q_2\sqrt{1-x_{3q}^2}\bigg),
\end{eqnarray}
\begin{eqnarray} \label{eq:c20}
\sin\phi_{3q}=&&\bigg(-p\sqrt{1-x_p^2}\sqrt{1-x_{\phi_p^2}}-\frac{1}{2}q\sqrt{1-x_q^2}\sqrt{1-x_{\phi_q}^2}\bigg)\nonumber
\\&&\bigg(q_2\sqrt{1-x_{3q}^2}\bigg).
\end{eqnarray}

For two-body current, i.e. the second term in Eq. (\ref{eq:c2}) one
can write as follows:
\begin{eqnarray} \label{eq:c21}
&&TBC=3\sum_{m'_2m'_3\nu'_2\nu'_3}\int_0^{\infty} dp'
p'^2\int_{-1}^1dx_{p'}\int_{-1}^1\frac{dx_{\phi_{p'}}}{\sqrt{1-x_{\phi_{p'}}^2}}\nonumber
\\&&\times\bigg\{F_{TBC}(x_{\phi_{p'}},\sqrt{1-x_{\phi_{p'}}^2})\nonumber \\&&+F_{TBC}(-x_{\phi_{p'}},-\sqrt{1-x_{\phi_{p'}}^2})\bigg\},
\end{eqnarray}
Where
\begin{eqnarray} \label{eq:c21.5}
&&F_{TBC}(x_{\phi_{p'}},\sqrt{1-x_{\phi_{p'}}^2})\nonumber
\\&&=j_{m_2m_3m'_2m'_3}^{\nu_2\nu_3\nu'_2\nu'_3}(q_{2b},x_{q_{2b}},x_{\phi_{2b}})\nonumber \\ &&\times\psi_{m_1m'_2m'_3}^{\nu_1\nu'_2\nu'_3}(p',q',x_{p'},x_{q'},x_{\phi_{p'}},x_{\phi_q})\nonumber
\\&& +j_{m_3m_1m'_2m'_3}^{\nu_3\nu_1\nu'_2\nu'_3}(q'_{2b},x_{q'_{2b}},x_{\phi'_{2b}})\nonumber \\ &&\times\psi_{m_2m'_2m'_3}^{\nu_2\nu'_2\nu'_3}(p',q',x_{p'},x_{q'},x_{\phi_{p'}},x_{\phi_q})\nonumber
\\&&
+j_{m_1m_2m'_2m'_3}^{\nu_1\nu_2\nu'_2\nu'_3}(q''_{2b},x_{q''_{2b}},x_{\phi''_{2b}})\nonumber
\\&&\times\psi_{m_3m'_2m'_3}^{\nu_3\nu'_2\nu'_3}(p',q',x_{p'},x_{q'},x_{\phi_{p'}},x_{\phi_q}),\;\;\;\;
\end{eqnarray}
 In term
of $p,q,x_p,x_q,x_{\phi_p},x_{\phi_q}$ we write the $j$'s argument
as follows:
\begin{eqnarray} \label{eq:c22}
q_{2b}&&=|\frac{1}{2}\textbf{Q}+\textbf{p}-\textbf{p}'|\nonumber
\\&&=(\frac{1}{4}Q^2+p^2+p'^2+pQx_p\nonumber \\&&-p'Qx_p'-pp'\cos\beta)^{\frac{1}{2}},
\end{eqnarray}
\begin{eqnarray} \label{eq:c23}
\cos\beta&&=x_px_p'+\sqrt{1-x_p^2}\sqrt{1-x_p'^2}\nonumber
\\&&\bigg(x_{\phi_{p'}}x_{\phi_p}+\sqrt{1-x_{\phi_{p'}}^2}\sqrt{1-x_{\phi_p}^2}\bigg),
\end{eqnarray}
\begin{eqnarray} \label{eq:c24}
x_{q_{2b}}=\frac{\frac{1}{2}Q+px_p-p'x_p'}{q_2},
\end{eqnarray}
\begin{eqnarray} \label{eq:c25}
x_{\phi_{2b}}=\cos\phi_{2b}=\frac{p\sqrt{1-x_p^2}x_{\phi_p}-p'\sqrt{1-x_p'^2}x_{\phi_{p'}}}{q_2\sqrt{1-x_{q_2}}},
\end{eqnarray}
\begin{eqnarray} \label{eq:c25.5}
\sin\phi_{2b}=&&\bigg(p\sqrt{1-x_p^2}\sqrt{1-x_{\phi_p}^2}-p'\sqrt{1-x_p'^2}\sqrt{1-x_{\phi_{p'}}^2}\bigg)\nonumber
\\&&/\bigg(q_2\sqrt{1-x_{q_2}}\bigg),
\end{eqnarray}
\begin{eqnarray} \label{eq:c26}
q'_{2b}&&=|\frac{1}{2}\textbf{Q}-\frac{1}{2}\textbf{p}-\frac{3}{4}\textbf{q}-\textbf{p}'|\nonumber
\\&&=(\frac{1}{4}Q^2+\frac{1}{4}p^2+\frac{9}{4}q^2+p'^2-\frac{1}{2}pQx_p-p'Qx_p'\nonumber
\\&&-\frac{3}{4}Qqx_q+pp'\cos\beta+\frac{3}{4}pq\cos\gamma'
\nonumber \\&&+\frac{3}{2}qp'\cos\beta')^{\frac{1}{2}},
\end{eqnarray}
\begin{eqnarray} \label{eq:c27}
\cos\beta'&&=x_{p'}x_q+\sqrt{1-x_{p'}^2}\sqrt{1-x_q^2}\nonumber
\\&&\bigg(x_{\phi}x_{\phi_{p'}}+\sqrt{1-x_{\phi}^2}\sqrt{1-x_{\phi_{p'}}^2}\bigg),
\end{eqnarray}
\begin{eqnarray} \label{eq:c28}
x_{q'_{2b}}=\frac{\frac{1}{2}Q-\frac{1}{2}px_p-\frac{3}{4}qx_q-p'x_{p'}}{q'_2},
\end{eqnarray}
\begin{eqnarray} \label{eq:c29}
x_{\phi'_{2b}}&&=\cos\phi'_{2b}\nonumber
\\&&=\bigg(-\frac{1}{2}p\sqrt{1-x_p^2}x_{\phi_p}\nonumber \\&&
-\frac{3}{4}q\sqrt{1-x_q^2}x_{\phi_q}-p'\sqrt{1-x_p'^2}x_{\phi_{p'}}\bigg)\nonumber
\\&&/\bigg(q'_2\sqrt{1-x_{q'}^2}\bigg),
\end{eqnarray}
\begin{eqnarray} \label{eq:c29.5}
\sin\phi'_{2b}&&=\bigg(-\frac{1}{2}p\sqrt{1-x_p^2}\sqrt{1-x_{\phi_p}^2}\nonumber
\\&&-\frac{3}{4}q\sqrt{1-x_q^2}\sqrt{1-x_{\phi_q}^2}-p'\sqrt{1-x_p'^2}\sqrt{1-x_{\phi_{p'}}^2}\bigg)
\nonumber \\&&/\bigg(q'_2\sqrt{1-x_{q'}^2}\bigg),
\end{eqnarray}
\begin{eqnarray} \label{eq:c30}
q''_{2b}&&=|\frac{1}{2}\textbf{Q}-\frac{1}{2}\textbf{p}+\frac{3}{4}\textbf{q}-\textbf{p}'|\nonumber
\\&&=(\frac{1}{4}Q^2+\frac{1}{4}p^2+\frac{9}{4}q^2+p'^2-\frac{1}{2}pQx_p-p'Qx_p'\nonumber
\\&&+\frac{3}{4}Qqx_q+pp'\cos\beta-\frac{3}{4}pq\cos\gamma\nonumber \\&&-\frac{3}{2}qp'\cos\beta')^{\frac{1}{2}},
\end{eqnarray}
\begin{eqnarray} \label{eq:c31}
x_{q''_{2b}}=\frac{\frac{1}{2}Q-\frac{1}{2}px_p+\frac{3}{4}qx_q-p'x_{p'}}{q''_2},
\end{eqnarray}
\begin{eqnarray} \label{eq:c32}
x_{\phi'_{2b}}&&=\cos\phi''_{2b}\nonumber
\\&&=\bigg(-\frac{1}{2}p\sqrt{1-x_p^2}x_{\phi_p}\nonumber \\&&+\frac{3}{4}q\sqrt{1-x_q^2}x_{\phi_q}-p'\sqrt{1-x_p'^2}x_{\phi_{p'}}\bigg)\nonumber \\&&/\bigg(q''_2\sqrt{1-x_{q''}^2}\bigg),
\end{eqnarray}
\begin{eqnarray} \label{eq:c32.5}
\sin\phi''_{2b}&&=\bigg(-\frac{1}{2}p\sqrt{1-x_p^2}\sqrt{1-x_{\phi_p}^2}\nonumber
\\&&+\frac{3}{4}q\sqrt{1-x_q^2}\sqrt{1-x_{\phi_q}^2}-p'\sqrt{1-x_p'^2}\sqrt{1-x_{\phi_{p'}}^2}\bigg)\nonumber \\&&/\bigg(q''_2\sqrt{1-x_{q''}^2}\bigg),
\end{eqnarray}
\begin{eqnarray}
q'=|\textbf{q}-\frac{1}{3}\textbf{Q}|=(q^2+\frac{1}{9}Q^2-\frac{2}{3}qQx_q)^{\frac{1}{2}},
\end{eqnarray}
\begin{eqnarray}
x_{q'}=\frac{qx_q-\frac{1}{3}Q}{q'}.
\end{eqnarray}
The third part of Eq. (\ref{eq:c2}) in term of the integration
variables and the six variables of the U function can be written as:
\begin{eqnarray} \label{eq:c33}
I&&=\frac{4}{q}\frac{1}{E-E_d-\frac{3q}{4m}}\sum_{m'_2m'_3\nu'_2\nu'_3}\int_0^{\infty}dp'\frac{p'}{E-\frac{3q^2}{4m}-\frac{p'^2}{m}}\nonumber
\\&&\int_{|\frac{q}{2}-p'|}^{\frac{q}{2}+p'}q'' q''\int_{-1}^1
dx_{q''}\int_{-1}^1\frac{dx_\phi''}{\sqrt{1-x_{\phi''}^2}}\nonumber
\\
&&\times\bigg\{F(x_{\phi''},\sqrt{1-x_{\phi''}^2})+F(-x_{\phi''},-\sqrt{1-x_{\phi''}^2})\bigg\}\nonumber
\\&&\times\delta(x''-x_0),\;\;\;
\end{eqnarray}
Where
\begin{eqnarray} \label{eq:c34}
F(x_{\phi''},\sqrt{1-x_{\phi''}^2})=\hat{t}^a(p,p',x_p,x_{p'},x_{\phi_p},x_{\phi_{p'}})
\nonumber \\ \times
U(p'',q'',x_{p''},x_{q''},x_{\phi_{p''},x_{\phi_{q''}}}),
\end{eqnarray}
\begin{eqnarray} \label{eq:c35}
x_{p'}=\frac{\frac{1}{2}qx_q+q''x_{q''}}{|\frac{1}{2}\textbf{q}+\textbf{q}''|},
\end{eqnarray}
\begin{eqnarray} \label{eq:c36}
|\frac{1}{2}\textbf{q}+\textbf{q}''|=(\frac{1}{4}q^2+q''^2+qq''x'')^{\frac{1}{2}},
\end{eqnarray}
\begin{eqnarray} \label{eq:c37}
&&x_{\phi_{p'}}=\cos\phi_{p'}\nonumber
\\&&=\frac{\frac{1}{2}q\sqrt{1-x_q^2}x_{\phi_q}+q''\sqrt{1-x_{q''}}x''_{\phi}}{|\frac{1}{2}\textbf{q}+\textbf{q}''|\sqrt{1-x_{p'}^2}},
\end{eqnarray}
\begin{eqnarray} \label{eq:c37.5}
\sin{\phi_{p'}}=&&\bigg(\frac{1}{2}q\sqrt{1-x_q^2}\sqrt{1-x_{\phi_q}^2}+q''\sqrt{1-x_{q''}^2}\sqrt{1-{x''}_{\phi}^2}\bigg)\nonumber
\\&&/\bigg(|\frac{1}{2}\textbf{q}+\textbf{q}''|\sqrt{1-x_{p'}^2}\bigg),
\end{eqnarray}
\begin{eqnarray} \label{eq:c38}
p''=\sqrt{p'^2-\frac{3{q'}^2}{4}+\frac{3q^2}{4}},
\end{eqnarray}
\begin{eqnarray} \label{eq:c39}
x_{p''}=\frac{-\frac{1}{2}q''x_{q''}-qx_{q}}{|\frac{1}{2}\textbf{q}''+\textbf{q}|},
\end{eqnarray}
\begin{eqnarray} \label{eq:c40}
|\textbf{q}+\frac{1}{2}\textbf{q}''|=(\frac{1}{4}q''^2+q'^2+qq''x'')^{\frac{1}{2}},
\end{eqnarray}
\begin{eqnarray} \label{eq:c41}
x''&&=x_qx_{q''}+\sqrt{1-x_q^2}\sqrt{1-x_{q''}^2}\nonumber
\\&&\bigg(x_{\phi''}x_{\phi_q}+\sqrt{1-x_{\phi''}^2}\sqrt{1-x_{\phi_q}^2}\bigg).
\end{eqnarray}
For delta function we have:
\begin{eqnarray} \label{eq:c43}
\delta(x''-x_0)=\delta[f(x_{q''})],
\end{eqnarray}
where:
\begin{eqnarray} \label{eq:c44}
f(x_{q''})=ax_{q''}+\sqrt{1-a^2}\sqrt{1-x_{q''}^2}c-x_0,
\end{eqnarray}
in the above equation, $a=x_q$, $c=x_{\phi_{qq''}}$ and
$x_0=\frac{p'^2-\frac{1}{4}q^2-{q''}^2}{qq''}$.

For handling the Eq. (\ref{eq:c43}) we have to use delta function
properties to change the argument from $x''$ to $x_{q''}$ . So we
need zero points of $f(x)$, i.e. $x_i$:
\begin{eqnarray} \label{eq:c45}
x_i=\frac{a^2x_0^2\pm\sqrt{\Delta}}{c^2(a^2-1)-a^2},
\end{eqnarray}
Where
\begin{eqnarray} \label{eq:c46}
\Delta=c^2(a^2-1)[c^2(a^2-1)+x_0^2-a^2].
\end{eqnarray}

\section{Two-body t-matrix, Deuteron and Triton wave functions in
free basis states} \label{app:tmatrix deuteron and triton} Two-body
calculations in the 3D approach are performed in Refs.
\cite{Fachruddin_PhRC62}-\cite{Fachruddin_PhRC63}. We briefly
introduce the two-body t-matrix in the momentum-helicity basis
states as:
\begin{eqnarray} \label{eq:39}
^{a\pi}\langle\textbf{p}\hat{p'}S\Lambda
T|t|\textbf{p}\hat{p}S\Lambda T\rangle^{\pi a}=t_{\Lambda'
\Lambda}^{\pi\,S\,T}(\textbf{p}',\textbf{p}),
\end{eqnarray}
Where the momentum-helicity basis states is defined by:
\begin{eqnarray} \label{eq:40}
|\textbf{p}\hat{p}S\Lambda T\rangle^{\pi
a}=(1-\eta_{\pi}(-1)^{S+T})|\textbf{p}\hat{p}S\Lambda\rangle_{\pi}|T\rangle,
\end{eqnarray}
where $S$ is spin, $\Lambda$ is its projection to $\hat{p}$ and $T$
is the isospin of the two-body system. If we choose the direction of
the vector $\textbf{p}$ to be in the $z$-direction we can write:
\begin{eqnarray} \label{eq:41}
t_{\Lambda'
\Lambda}^{\pi\,S\,T}(\textbf{p}',\textbf{p})=e^{i\lambda\phi_{p'}}t_{\Lambda'
\Lambda}^{\pi\,S\,T}(p',p,x_{p'}),
\end{eqnarray}
and the Lippmann-Schwinger equation in the momentum-helicity basis
states for any energy is written as:
\begin{eqnarray} \label{eq:42}
&&t_{\Lambda' \Lambda}^{\pi\,S\,T}(p',p,x_{p'})=V_{\Lambda'
\Lambda}^{\pi\,S\,T}(p',p,x_{p'})\nonumber
\\&&+\frac{1}{4}\sum_{\Lambda''}\int_0^\infty
q''q''^2\int_{-1}^1dx''v_{\Lambda'
\Lambda''}^{\pi\,S\,T,\Lambda}(p',p'',x_{p'},x'')\nonumber
\\&&\times G_0^+(E_p)t_{\Lambda''
\Lambda}^{\pi\,S\,T}(p'',p,x''),
\end{eqnarray}
Where
\begin{eqnarray} \label{eq:43}
&&v_{\Lambda'
\Lambda''}^{\pi\,S\,T,\Lambda}(p',p'',x_{p'},x'')\nonumber
\\&&=\int_0^{2\pi}d\phi''e^{-i\Lambda(\phi'-\phi'')}V_{\Lambda'
\Lambda''}^{\pi\,S\,T}(\textbf{p}',\textbf{p}'').
\end{eqnarray}
The relation between t-matrix in the momentum-helicity basis states
and those in the free basis states is:
\begin{eqnarray} \label{eq:44}
 &&^a\langle
\textbf{p}'m'_2m'_3\nu'_2\nu'_3|t|\textbf{p}m_2m_3\nu_2\nu_3\rangle^a
\nonumber
\\&&=\frac{1}{4}\delta_{(\nu_2+\nu_3),(\nu'_2+\nu'_3)}e^{i(\Lambda_0\phi_p-\Lambda'_0\phi_{p'})}\sum_{\Lambda S T}(1-\eta_{\pi}(-1)^{S+T})\nonumber \\
&&\times
C(\frac{1}{2}\frac{1}{2}T,\nu_2\nu_3)C(\frac{1}{2}\frac{1}{2}T,\nu'_2\nu'_3)C(\frac{1}{2}\frac{1}{2}S,m_2m_3\Lambda_0)\nonumber
\\ &&\times C(\frac{1}{2}\frac{1}{2}S,m'_2m'_3\Lambda'_0)\sum_{\Lambda\Lambda'}d_{\lambda_0\Lambda}^S(x_p)d_{\Lambda'_0\Lambda'}^S(x_{p'})\nonumber \\&&t_{\Lambda'
\Lambda}^{\pi\,S\,T}(\textbf{p}',\textbf{p}),
\end{eqnarray}
rotation matrix element, $d_{\lambda'\Lambda}^S(x_p)$ is introduced
in (\ref{eq:b7}). The deuteron wave function in the free basis
states is also related to the momentum-helicity basis states:
\begin{eqnarray} \label{eq:50}
 &&\langle\psi_d^{M_d}|\textbf{p}m_2m_3\nu_2\nu_3\rangle^a\nonumber\\
&&
=\frac{1}{2}\bigg[\frac{1}{2}\phi_0^{M_d}(\textbf{p})d_{\Lambda_00}^1(x_p)-\frac{1}{2}\phi_0^{M_d}(-\textbf{p})d_{\Lambda_00}^1(x_p)\nonumber
 \\ &&+\phi_1^{M_d}(\textbf{p})d_{\Lambda_01}^1(x_p)-\phi_0^{M_d}(-\textbf{p})d_{\Lambda_0-1}^1(x_p)\bigg]\nonumber
 \\ &&\times C(\frac{1}{2}\frac{1}{2}1,m_2m_3\Lambda_0)C(\frac{1}{2}\frac{1}{2}0,\nu_2\nu_30)\nonumber
\\&&e^{i\Lambda_0\phi_q}d_{\Lambda_0\Lambda}^1(x_q).
\end{eqnarray}
Where $M_d$ is spin projection of the deuteron, we have summed over
$\Lambda=-1,0,1$, and we have used this fact that the terms with
$\Lambda=-1$ and $\Lambda=1$ are the same. Also we define:
\begin{eqnarray} \label{eq:48}
\phi_{\Lambda}^{M_d}(\textbf{q})=\,^{1a}\langle\textbf{q}\hat{q}1\Lambda;0|\psi_d^{M_d}\rangle=\phi_{\Lambda}^{M_d}(q,x_q)e^{iM_d\phi_q},
\end{eqnarray}
the deuteron wave function in the momentum-helicity basis states is
calculated using eigenvalue equation:
\begin{eqnarray} \label{eq:51}
&&(\frac{q^2}{m}-E_d)\phi_{\Lambda}^{M_d}(q,x_q)+\frac{1}{2}\int_{-\infty}^{\infty}dq
q^2\nonumber
\\&&\int_{-1}^{1}dx'v_{\Lambda1}^{110,M_d}(q,q',x_q,x')\phi_{1}^{M_d}(q',x')+\frac{1}{4}\int_{-\infty}^{\infty}dq
q^2\nonumber
\\&&\int_{-1}^{1}dx'v_{\Lambda0}^{110,M_d}(q,q',x_q,x')\phi_{0}^{M_d}(q',x')=0.
\end{eqnarray}

The triton wave function is calculated by applying the formalism of
Ref.\cite{Bayegan_PhRC77}. In this formalism the triton wave
function has been evaluated in the basis states:
\begin{eqnarray} \label{eq:52}
\langle\textbf{p}\textbf{q}\alpha|\Psi\rangle=\langle\textbf{p}\textbf{q}(s_{23},\frac{1}{2})SM_S(t_{23},\frac{1}{2})TM_T|\Psi\rangle,
\end{eqnarray}
 In the above equation $s_{23}$ is total spin of the subsystem $S$
 and $M_S$ are the total spin of the three particles and
 its projection along the z axis respectively. The same explanation is used for the isospin.

The triton wave function in Eq. (\ref{eq:52}) is reproduced using
AV18 potential and is related to the one in the free basis states as
follows:
\begin{eqnarray} \label{eq:53}
\langle\textbf{p}\textbf{q}m_1m_2m_3\nu_1\nu_2\nu_3|\Psi\rangle&&=\sum_{\alpha}
\langle
m_1m_2m_3\nu_1\nu_2\nu_3|\alpha\rangle\langle\textbf{p}\textbf{q}\alpha|\Psi\rangle\nonumber
\\&&=\sum_{\alpha}g_{\gamma\alpha}\langle\textbf{p}\textbf{q}\alpha|\Psi\rangle,
\end{eqnarray}
$|\alpha\rangle$ and $|\gamma\rangle$ are spin and isospin states
that are defined by  Eqs. (\ref{eq:54}) and (\ref{eq:54p}) and the
Clebsch-Gordan coefficients, $g_{\gamma\alpha}$, are introduced in
Ref. \cite{Bayegan_PhRC77}.

\begin{eqnarray} \label{eq:54}
&&|\alpha\rangle=|(s_{23},\frac{1}{2})SM_S(t_{23},\frac{1}{2})TM_T\rangle,
\end{eqnarray}
\begin{eqnarray} \label{eq:54p}
&&|\gamma\rangle=|m_1m_2m_3\nu_1\nu_2\nu_3\rangle.
\end{eqnarray}

\section{Extraction of Deuteron singularity}\label{app:Deuteron}

To extract singularity of the two-body t-matrix in deuteron binding
energy we need to evaluate the following term in the
momentum-helicity basis states.

\begin{eqnarray} \label{eq:d1}
&&\lim_{E\rightarrow E_d}(E-E_d)t_{\Lambda'\Lambda}^{110}\nonumber
\\&&=\,^{1a}\langle\textbf{p}'\hat{p'}1\Lambda';0|V|\Psi_d^{M_d}\rangle\langle\Psi_d^{M_d}|V|\textbf{p}\hat{p}1\Lambda;0\rangle^{1a},
\end{eqnarray}
two terms on the right hand side of the above equation evaluate
separately.
\begin{eqnarray} \label{eq:d2}
&&^{1a}\langle\textbf{p}'\hat{p'}1\Lambda';0|V|\Psi_d^{M_d}\rangle\nonumber
\\&&=\frac{1}{4} \sum_{\Lambda''}\int
d\textbf{p}''\,^{1a}\langle\textbf{p}'\hat{p'}1\Lambda';0|V|\textbf{p}''\hat{p''}1\Lambda'';0\rangle^{1a}\nonumber
\\&&\times\,^{1a}\langle\textbf{p}''\hat{p''}1\Lambda'';0|\Psi_d^{M_d}\rangle\nonumber
\\&&=\frac{1}{4}\sum_{\Lambda''}\int
d\textbf{p}''V_{\Lambda'\Lambda''}^{110}(\textbf{p}',\textbf{p}'')\Phi_{\Lambda''}^{M_d}(\textbf{p}''),
\end{eqnarray}
and
\begin{eqnarray} \label{eq:d3}
&&\langle\Psi^{M_d}|V|\textbf{p}\hat{p}1\Lambda0\rangle^{1a}\nonumber
\\&&=\frac{1}{4}\sum_{\Lambda''}\int
d\textbf{p}''\langle\Psi^{M_d}|V|\textbf{p}''\hat{p''}1\Lambda0\rangle^{1a}\nonumber
\\&&\times\,^{1a}\langle\textbf{p}'';\hat{p''}1\Lambda0|V|\textbf{p}\hat{p}1\Lambda0\rangle^{1a}\nonumber
\\&&=\frac{1}{4}\sum_{\Lambda''}\int
d\textbf{p}''V_{\Lambda''\Lambda}^{110}(\textbf{p}'',\textbf{p})\Phi_{\Lambda''}^{*M_d}(\textbf{p}'').
\end{eqnarray}

By multiplying these two terms and considering the momentum vector
$\textbf{p}$ to be in z direction,we can extract the azimuthal angle
parts of the potential and the deuteron wave function by applying
Eqs. (\ref{eq:41}) and (\ref{eq:48}). We then multiply
$e^{-i(\Lambda-M_d)\Phi'}$ to the both sides of Eq. (\ref{eq:d1})and
we derive the following equation using Eq. (\ref{eq:43}):
\begin{eqnarray} \label{eq:d7}
&&\lim_{E\rightarrow E_d}(E-E_d)t_{\Lambda'\Lambda}^{110}(p',p,x')\nonumber \\
&&=\bigg\{\frac{1}{4}\int
p''^2dp''dx''\bigg(v_{\Lambda'0}^{110,M_d}(p',x',p'',x'')\Phi_{0}^{M_d}(p'',x'')\nonumber
\\
&&+2v_{\Lambda'1}^{110,M_d}(p',x',p'',x'')\Phi_{1}^{M_d}(p'',x'')\bigg)\bigg\}\nonumber
\\ &&\times\bigg\{\frac{1}{4}\int
p''^2dp''dx''\bigg(V_{0\Lambda}^{110}(p'',p,x'')\Phi_{0}^{*M_d}(p'',x'')\nonumber
\\
&&+2V_{1\Lambda}^{110}(p'',p,x'')\Phi_{1}^{*M_d}(p'',x'')\bigg)\int_{0}^{2\pi}
d\Phi''e^{i(\Lambda-M_d)(\Phi''-\Phi')}\bigg\}\nonumber \\
&&=\bigg\{\frac{1}{4}\int
p''^2dp''dx''\bigg(v_{\Lambda'0}^{110,\Lambda}(p',x',p'',x'')\Phi_{0}^{\Lambda}(p'',x'')\nonumber
\\
&&+2v_{\Lambda'1}^{110\Lambda}(p',x',p'',x'')\Phi_{1}^{\Lambda}(p'',x'')\bigg)e^{-iM_d(\Phi'-\Phi''}\bigg\}\nonumber
\\ &&\times\bigg\{\frac{1}{4}\int
p''^2dp''dx''\bigg(V_{0\Lambda}^{110}(p'',p,x'')\Phi_{0}^{*\Lambda}(p'',x'')\nonumber
\\
&&+2V_{1\Lambda}^{110}(p'',p,x'')\Phi_{1}^{*\Lambda}(p'',x'')\bigg)\bigg\}.
\end{eqnarray}
In above equation we have used this equality:
\begin{eqnarray} \label{eq:d8}
\int_{0}^{2\pi}
d\Phi''e^{i(\Lambda-M_d)(\Phi''-\Phi')}=2\pi\delta_{\Lambda M_d}
\end{eqnarray}

%

%



\begin{thebibliography}{}
%
\bibitem{Faddeev_ThF39}
L. D. Faddeev, Zh. Eksp. Theor. Fiz. \textbf{39}, 1459 (1960).
\bibitem{Alt_NPB2}
E. O. Alt, P. Grassberger, W. Sandhas, Nucl. Phys. \textbf{B2}, 167
(1967).

\bibitem{Lehman_PhRL23}
D. R. Lehman, Phys. Rev. Lett. \textbf{23}, 1339 (1969).
\bibitem{Barbour_PhRL19}
I. R. Barbour, A.C. Phillips, Phys. Rev. Lett. \textbf{19}, 1388
(1967).
\bibitem{Gibson_PhRC11}
B. F. Gibson and D. R. Lehman, Phys. Rev. \textbf{C 11}, 29 (1975).
\bibitem{Carlson_PhRC36}
J. Carlson, Phys. Rev. \textbf{C 36}, 2026 (1987).
\bibitem{Efros_PhLB338}
V. D. Efros, W. Leidemann, G. Orlandini, Phys. Lett. \textbf{B 338},
130 (1994).
\bibitem{Viviani_PhRC61}
M. Viviani, A. Kievsky, L. E. Marcucci, S. Rosati, R. Schiavilla,
Phys. Rev. \textbf{C 61}, 064001 (2000).
\bibitem{golak}
J.Golak et al. ,  Phys.Rept. \textbf{415}, 89 (2005).
\bibitem{rice}
R. A. Rice, Y. E. Kim, Few-Body Syst. \textbf{14}, 127 (1993).
\bibitem{Fachruddin_PhRC62}
I. Fachruddin, Ch. Elster, W. Gl\"{o}ckle, Phys. Rev. \textbf{C 62},
044002 (2000).
\bibitem{Fachruddin_PhRC63}
I. Fachruddin, Ch. Elster, W. Gl\"{o}ckle, Phys. Rev. \textbf{C 63},
054003 (2001).
\bibitem{hadizadeh-Eur.Phys}
M. R. Hadizadeh and S. Bayegan, Eur. Phys. J. \textbf{A 36}
201(2008).
\bibitem{Bayegan-Prog.Theor.Phys}
S. Bayegan, M. R. Hadizadeh, and W. Gl¨ockle, Prog. Theor. Phys.
\textbf{120} 887 (2008).
\bibitem{Bayegan-nucl814}
S. Bayegan, M. Harzchi and M. R. Hadizadeh, Nucl. Phys. \textbf{A
814} 21(2008).
\bibitem{Bayegan_PhRC77}
S. Bayegan, M. R. Hadizadeh, and M. Harzchi, Phys. Rev. \textbf{C
77}, 064005 (2008).
\bibitem{Bayegan_PhRC79}
S. Bayegan, M. A. Shalchi, M. R. Hadizadeh, Phys. Rev. \textbf{C
79}, 057001 (2009).
\bibitem{Bayegan-nucl832}
S. Bayegan, M. Harzchi and M. A. Shalchi, Nucl. Phys. \textbf{A 832}
1(2010).
\bibitem{harzchi-nucla46}
M. Harzchi, S. Bayegan, Eur. Phys. J. \textbf{A 46} 271 (2010).
\bibitem{Hadizadeh-Phys. Revc83}
M. R. Hadizadeh, Lauro Tomio, S. Bayegan, Phys. Rev. \textbf{C 83},
054004 (2011).
\bibitem{Elster_FbS45}
Ch. Elster, W. Gl\"{o}ckle, H. Witala, Few Body Syst. \textbf{45}, 1
(2009).
\bibitem{H. Liu}
H. Liu, Ph.D thesis. Ohio University, USA, (2005).
\bibitem{AV18}
R. B. Wiringa, V. G. J. Stoks, R. Schiavilla, Phys. Rev. \textbf{C
51}, 38 (1995).
\bibitem{spline}
D. H¨uber, H. Witala, A. Nogga, W. Gl¨ockle and H. Kamada, Few-Body
Systems \textbf{22}, 107 (1997).
\bibitem{riska}
D.O. Riska, Phys. Scr. 31 107(1985).
\bibitem{seigert}
A.J.F. Siegert, Phys. Rev. \textbf{52} 787 (1937).
\bibitem{Faul_PhRC24}
D. D. Faul, B. L. Berman, P. Meyer, D. L. Olson, Phys. Rev.
\textbf{C 24}, 849 (1981).
\bibitem{benchmark}
J. Golak et al. Nucl. Phys \textbf{A 707}, 365 (2002).
\bibitem{Skopik_PhRC24}
D. M. Skopik, D. H. Beck, J. Asai, J. J. Murphy II, Phys. Rev.
\textbf{C 24}, 1791 (1981).
\bibitem{Kosiek_PhL21}
R. Kosiek, D. M\"{u}ller, R. Pfeiffer, O. Merwitz, Phys. Lett.
\textbf{21}, 199 (1966).
\end{thebibliography}
\end{document}